\documentclass[11pt, a4paper]{article}
\pdfoutput=1
\usepackage{jheppub}

\usepackage[english,greek]{babel}
\usepackage{verbatim}

\usepackage{color}

\usepackage{amsmath}
\usepackage{amssymb}
\usepackage{amsthm}
\usepackage{amsfonts}
\usepackage{mathtools}
\usepackage{simplewick}

\usepackage{endnotes}
\usepackage{chngcntr}
\usepackage[retainorgcmds]{IEEEtrantools}

\usepackage[]{mciteplus}
\mciteSetBstSublistMode{n}
\mciteSetMidEndSepPunct{\quad$\bullet$\quad}{.}{\relax}
\usepackage{graphicx}

\counterwithin{equation}{section}
\newtheorem{proposition}{$\blacksquare$}[subsection]
\newtheorem{theorem}{$\blacksquare$ Theorem}

\title{Stringy Membranes in AdS/CFT.}
\author[a]{Minos Axenides,}
\author[b,1]{Emmanuel Floratos\note{On leave from the Department of Physics, National and Kapodistrian University of Athens, Zografou Campus, 157 84, Athens, Greece.}}
\author[a,c]{and Georgios Linardopoulos}
\affiliation[a]{Institute of Nuclear and Particle Physics, N.C.S.R. "Demokritos",\\
153 10, Agia Paraskevi, Greece. \\}
\affiliation[b]{Department of Physics, Theory Division, CERN, \\ CH–1211 Geneva 23, Switzerland. \\}
\affiliation[c]{Department of Physics, National and Kapodistrian University of Athens,\\
Zografou Campus, 157 84, Athens, Greece. \\}
\emailAdd{axenides@inp.demokritos.gr}
\emailAdd{mflorato@phys.uoa.gr}
\emailAdd{glinard@inp.demokritos.gr}
\abstract{We study membrane configurations in AdS$_{7/4} \times S^{4/7}$. The membranes are wrapped around the compact manifold $S^{4/7}$ and are dynamically equivalent to bosonic strings in AdS$_5$. We thus conveniently identify them as "Stringy Membranes". For the case of $\text{AdS}_7\times S^4$, their construction is carried out by embedding the Polyakov action for classical bosonic strings in $\text{AdS}_5$, into the corresponding membrane action. Therefore, every string configuration in AdS$_5$ can be realized by an appropriately chosen stringy membrane in $\text{AdS}_7\times S^4$. We discuss the possibility of this being also the case for stringy membranes in AdS$_4 \times S^7 / \mathbb{Z}^k$ ($k \geq 1$). By performing a stability analysis to the constructed solutions, we find that the (membrane) fluctuations along their transverse directions are organized in multiple Lam\'{e} stability bands and gaps in the space of parameters of the configurations. In this membrane picture, strings exhibit a single band/gap structure.}
\keywords{M-Theory, p-branes, AdS/CFT Correspondence, Gauge-gravity correspondence.}
\arxivnumber{1306.0220}
\begin{document}
\selectlanguage{english}
\maketitle\flushbottom
\normalsize
\section[Introduction.]{Introduction.}
Extended objects such as strings and membranes have played an important role in our understanding of fundamental interactions. Membranes first appeared in the early attempt of Dirac \cite{Dirac62} to model an electron by a charged closed membrane and underlay the development of hadronic bag models \cite{ChodosJaffeJohnsonThornWeisskopf74a, BardeenChanowitzDrellWeinsteinYan75} (for a review see \cite{HasenfratzKuti77}). The emergence of Yang-Mills theory as the conceptual foundation of the standard model of strong and electroweak interactions, brought about the still unresolved puzzle of color confinement in QCD. That strings model successfully the almost linear Regge trajectories, does not belittle the virtues of bosonic membranes as effective bags of the QCD vacuum, which also recover successfully the string limit. At this point we should also mention the striking equivalence between regularized spherical bosonic membranes and SU$(\infty)$ classical Yang-Mills theory, first observed by Goldstone and Hoppe \cite{Hoppe82}.\footnote{This is due to the fact that the group of area-preserving diffeomorphisms (SDiff) of spherical membranes can be approximated by SU$(N)$ \cite{FloratosIliopoulos88, AntoniadisDitsasFloratosIliopoulos88, FloratosIliopoulosTiktopoulos89}. It also holds true for surfaces of any genus (see e.g. \cite{Taylor01} and references therein).} Moreover, let us also stress the analogies between the topological structure of Yang-Mills theory (self-duality) and closed bosonic membranes \cite{BiranFloratosSavvidy87, FloratosLeontaris89b}. \\[6pt]
Besides offering a model for the description of elementary particles, quantum relativistic membranes became widely known as 2-dimensional generalizations of strings \cite{CollinsTucker76}. In contrast to strings however and due to their lacking a coupling constant, membranes are notoriously non-perturbative objects. As a consequence, systematic perturbative methods have not been developed for membranes, making them less attractive than strings as fundamental building blocks of matter. Indeed, the first superstring revolution underlined the prime role of superstrings as fundamental constituents. Despite that, the M-theory revolution \cite{Witten95a, Townsend95a} and later the "Matrix Theory Conjecture" \cite{BFSS97} (see \cite{BigattiSusskind97} for a review), paved the way for a more "democratic" framework \cite{Townsend95b} into which strings, membranes and p-branes of various dimensionalities coexist \cite{Duff96}. Indeed, the web of 10-dimensional string theories finds a unifying platform in a single 11-dimensional M-theory whose long-wavelength limit is just 11-dimensional supergravity.\\[6pt]
In all of these developments, flat Minkowski spacetimes provided the majority of backgrounds for the study of membranes \cite{KikkawaYamasaki86, HoppeNicolai87, BarsPopeSezgin87, BergshoeffSezginTownsend88}. The spacetimes AdS$_{4/5/7} \times S^{7/5/4}$ were known as (maximally supersymmetric) vacua of ten and eleven dimensional supergravity \cite{FreundRubin80, PilchvanNieuwenhuizenTownsend84} but they were only rarely used as membrane backgrounds \cite{Sezgin98}, despite furnishing them with exceptional features such as massless excitations and a discrete spectrum.\footnote{Owing to the periodicity of the temporal coordinate.} Moreover, the "membrane at the end of the universe" \cite{BergshoeffDuffPopeSezgin87, BergshoeffSezginTanii88, BergshoeffDuffPopeSezgin89, DuffPopeSezgin89}, as the membrane at the boundary of AdS$_4 \times S^7$ spacetime came to be known, seemed to give rise to an $OSp\left(8|4\right)$ superconformal field theory. The 1997 AdS/CFT correspondence of Maldacena \cite{Maldacena97, GubserKlebanovPolyakov98, Witten98a} (see \cite{MAGOO99, DHokerFreedman02} for reviews) grouped many of the deep ideas that were present in 't Hooft's large-N expansion \cite{Hooft74}, the holographic principle \cite{Hooft93, Susskind95} or the geometrization of RG flow (see e.g. \cite{Tseytlin88, CallanThorlacius89}) into a tractable model of gauge/gravity duality. \nocite{BergesTetradisWetterich00} \\[6pt]
According to the standard dictionary of the AdS/CFT correspondence, the energy of a state in the bulk of anti-de Sitter space, equals the scaling dimension of its dual CFT operator. In 2002, Gubser, Klebanov and Polyakov (GKP) \cite{GubserKlebanovPolyakov02} introduced a method for the calculation of the anomalous dimensions of certain local, gauge-invariant operators of $\mathcal{N} = 4$ super Yang-Mills theory at strong 't Hooft coupling, a regime which is classically inaccessible by ordinary perturbation techniques. Their method consists in studying closed strings that spin inside $AdS_5 \times S^5$ and in calculating their energy in terms of their other conserved charges, such as their spin or angular momentum, semiclassically. To every string state is then assigned an operator of $\mathcal{N} = 4$, SYM, the bare dimension of which is some function of its spin and $SU(4)$ quantum numbers.\footnote{$SU(4) \cong SO(6)$ is a bosonic subgroup of $PSU(2,2|4)$, the full symmetry supergroup of $\mathcal{N} = 4$, SYM and IIB string theory on $AdS_5 \times S^5$.} As the dimensions of the operators typically receive quantum corrections, their renormalized values (anomalous dimensions) at strong coupling are expected to coincide with the corresponding energies of their dual bulk states, as obtained by the semiclassical treatment of GKP.\\[6pt]
The GKP string serves as a benchmark of the Maldacena conjecture because it compares the spectra on both sides of the correspondence \cite{MAGOO99, DHokerFreedman02}, beyond BPS or nearly BPS (BMN) states. The proposal provoked a flurry of research activity (see e.g. \cite{Beisertetal12}). In addition to the AdS$_5$/CFT$_4$ proposal, the role of M-theory in AdS/CFT has been investigated in $AdS_7 \times S^4$ \cite{AharonyBerkoozKachruSeibergSilverstein97, Douglas10, LambertPapageorgakisSchmidtSommerfeld12} and $AdS_4 \times S^7/\mathbb{Z}_k$ backgrounds \cite{BaggerLambert07, ABJM08, BaggerLambertMukhiPapageorgakis12}. However, the existence of precise formulations of boundary theories for the AdS$_4$ class of backgrounds has no match with the AdS$_7$ ones. In the latter case potential interest arises through the work of Witten \cite{Witten98b} who showed that a model for large-N QCD$_4$ can be obtained by toroidally compactifying the CFT dual of M-theory on AdS$_7 \times S^4$.\nocite{GursoyKiritsisMazzantiNitti08} In this sense, the investigation of membrane solutions in these backgrounds gains in importance \cite{SezginSundell02, AlishahihaGhasemkhani02, AlishahihaMosaffa02, HartnollNunez02, HoppeTheisen04, BruguesRojoRusso04, ArnlindHoppeTheisen04, Bozhilov03, Bozhilov05, Bozhilov07a, AhnBozhilov08, KimKimLee10}. \\[6pt]
In our present work we construct membrane configurations in stringy disguise, which we conveniently call "stringy membranes". Their essential property is that they are partially wrapped around a compact dimension and reproduce the action, equations of motion and conserved charges of a string.\footnote{An interesting, yet questionable by many (since nonlinear sigma models in more than two dimensions are not renormalizable by power counting), application of wrapping is the semiclassical quantization of an 11-dimensional supermembrane that is wrapped around a torus \cite{DIPSS88}.} The bonus of this is twofold: firstly, at the level of classical quadratic fluctuations around stringy membrane solutions, we reveal the existence of an infinite set of purely membrane modes, in addition to the expected purely stringy ones. Secondly, just as the AdS$_5$/CFT$_4$ parameter matching affords to strings in the bulk of AdS an effective string tension $\sqrt{\lambda}$, our stringy membranes are similarly endowed with an effective tension $\sqrt{\lambda'} = R \, \sqrt{\lambda} / g_s \, \ell_s$.\footnote{R is the wrapping radius, $g_s$ the string coupling constant and $\ell_s$ the fundamental string length.}\\[6pt]
In order to construct configurations with the above properties, we embed the bosonic Polyakov action for strings in AdS$_5$ into the AdS$_7 \times S^4$ membrane action. We demonstrate that every AdS$_5$ string solution corresponds to a properly constructed membrane of AdS$_7 \times S^4$ and every AdS$_4 \subset$ AdS$_5$ string solution can be written as a membrane of AdS$_4 \times S^7/\mathbb{Z}_k$. An advantage of this construction can be seen through the quadratic fluctuation analysis around our specific stringy membrane solutions, which we perform in detail. We find that an independent subset of fluctuations, which is transverse to the direction of the stringy membrane, admits a Lam\'{e} multi-band/multi-gap structure, which is characteristic of their membrane nature. In our fluctuation analysis, string excitations are represented by single-band/single-gap configurations, suggesting that our AdS$_7$ membranes are collective excitations of their AdS$_5$ stringy counterparts. \\[6pt]
Our paper is organized as follows. We begin in section \ref{SpinningMembranes} with a brief reminder of the equations that determine the motion of a bosonic membrane in $AdS_7 \times S^4$. In section \ref{Strings&Membranes} we demonstrate how some simple $AdS_7 \times S^4$ membrane ans\"{a}tze reproduce the action and equations of motion of the following two spinning string configurations of \cite{GubserKlebanovPolyakov02}: (I) the $AdS_3$ closed \& folded string and (II) the string that pulsates in $AdS_5$. This is not a mere coincidence and we then proceed to prove (in conformity with \cite{DuffHoweInamiStelle87}) that all bosonic string ans\"{a}tze in $AdS_5$ that are consistent with the conformal gauge, can be generated by appropriate membrane ans\"{a}tze in  $AdS_7 \times S^4$. The extension of these considerations to $AdS_4 \times S^7/\mathbb{Z}_k$ is discussed in sections \ref{AdS4xS7} and \ref{AdS4xS7/Zk}. The stability of our solitons is examined in section \ref{Fluctuations}. We discuss our results in section \ref{Discussion}. In appendix \ref{Appendix:StringActions} we revisit the GKP string configurations (I) and (II) that we use in our paper and in appendix \ref{LameAppendix} we briefly discuss Lam\'{e}'s equation.
\newpage
\section[Spinning Membranes in $AdS_7 \times S^4$.]{Spinning Membranes in $AdS_7 \times S^4$. \label{SpinningMembranes}}

The bosonic part of the (Howe-Tucker-) Polyakov action \cite{HoweTucker77} for a membrane in D spacetime dimensions, in the presence of a Wess-Zumino flux term is: \\[12pt]
\small\begin{equation}
\mathcal{S} = - \frac{T_2}{2} \int d^3\sigma \left\{\sqrt{-\gamma} \left(\gamma^{ab} h_{ab} - 1\right) + 12 \, \dot{X}^m \, \partial_\sigma X^n \, \partial_\delta X^p \, A_{mnp}(X)\right\}, \quad T_2 \equiv \frac{1}{(2\pi)^2 \ell_P^3}, \label{Polyakov}
\end{equation} \\[6pt] \normalsize
where $\ell_P$ is the Planck length of D-dimensional spacetime, $X_m$ the spacetime coordinates, and $\sigma_a = \left\{\tau, \; \sigma, \; \delta\right\}$ are the membrane/world-volume coordinates ($\sigma, \delta \in \left[0,2\pi\right)$). On the other hand $A_{mnp}\left(X\right)$ is an antisymmetric 3-form tensor field, $g_{mn}\left(X\right)$ is the spacetime metric, $\gamma_{ab}$ the membrane/world-volume (auxiliary) metric and $h_{ab}$ is its induced metric on the membrane world-volume (pull-back): \\[6pt]
\begin{equation}
h_{ab} \equiv \partial_a X^m \partial_b X^n \, g_{mn}\left(X\right) = \gamma_{ab}, \quad h \equiv \det{h_{ab}},
\end{equation} \\
where $h_{ab} = \gamma_{ab}$ is the equation of motion that is obtained by varying action \ref{Polyakov} w.r.t. the auxiliary metric $\gamma_{ab}$. An especially convenient gauge choice is the following: \\[6pt]
\begin{equation}
\gamma_{00} = h_{00} = - \det h_{ij}\,, \qquad \gamma_{0i} = h_{0i} = 0\,, \qquad \gamma_{ij} = h_{ij}\,, \qquad i\,,j = 1, 2. \label{MembraneGauge}
\end{equation} \\[6pt]
The Polyakov action \ref{Polyakov} then becomes: \\[12pt]
\small\begin{equation}
\mathcal{S} = \frac{T_2}{2} \int d^3\sigma \left\{g_{mn} \dot{X}^m \dot{X}^n - \frac{1}{2} g_{mn} g_{pq}\{X^m, X^p\} \{X^n, X^q\} - 12 \, A_{mnp} \dot{X}^m \, \partial_\sigma X^n \, \partial_\delta X^p \right\}, \label{Polyakov1}
\end{equation} \\[6pt]\normalsize
where the Poisson bracket, $\{\;,\;\}$ is defined as: \\[6pt]
\begin{equation}
\left\{f\,,\,g\right\} \equiv \partial_\sigma f \; \partial_\delta g - \partial_\delta g \; \partial_\sigma f.
\end{equation} \\
The constraints that follow from fixing the gauge \ref{MembraneGauge} are:\\
\begin{IEEEeqnarray}{l}
\gamma_{00} = - \det h_{ij} \Rightarrow \; g_{mn} \dot{X}^m \dot{X}^n + \frac{1}{2} g_{mn} g_{pq}\{X^m, X^p\} \{X^n, X^q\} = 0 \label{MembraneConstraints1}\\[6pt]
\gamma_{0i} = g_{mn} \dot{X}^m \partial_i X^n = 0 \Rightarrow \left\{g_{mn} \, \dot{X}^m, X^n\right\} = 0. \label{MembraneConstraints2}
\end{IEEEeqnarray} \\[6pt]
Let us now briefly consider the general motion of a classical, uncharged (no WZ term) bosonic membrane in $AdS_7\times S^4$, as described in the global coordinate system of $AdS_7\times S^4$ (for AdS$_7 \times S^4$, it's $\ell = 2R$. Setting $R = 1$ implies that $\ell = 2$):\footnote{Our results are also valid for general R and $\ell$, proviso $\delta \mapsto \delta / R$ and $\delta \in \left[0,\,2 \pi R\right)$.}\\
\begin{IEEEeqnarray}{lcll}
Y_0 + i \, Y_7 = 2 \cosh\rho \, e^{it} \quad && X_1 + i & X_2 = \cos\overline{\theta}_1 \, e^{i \overline{\phi}_1} \nonumber \\[6pt]
Y_1 + i \, Y_2 = 2 \sinh\rho \cos\theta_1 \, e^{i\phi_1} & \quad \& \quad & X_3 + i & X_4 = \sin\overline{\theta}_1 \cos\overline{\theta}_2 \, e^{i \overline{\phi}_2} \\[6pt]
Y_3 + i \, Y_4 = 2 \sinh\rho \sin\theta_1 \cos\theta_2 \, e^{i\phi_2} &&& X_5 = \sin\overline{\theta}_1 \sin\overline{\theta}_2 \nonumber \\[6pt]
Y_5 + i \, Y_6 = 2 \sinh\rho \sin\theta_1 \sin\theta_2 \, e^{i\phi_3}, \nonumber
\end{IEEEeqnarray} \\
where $Y^\mu$ and $X^i$ are the embedding coordinates of AdS$_7 \times S^4$ (see \ref{EmbeddingCoordinates1}--\ref{EmbeddingCoordinates2}) and $\rho \geq 0, \,t \in \left[0, \, 2\pi\right),\footnote{Time periodicity is customarily avoided in all relevant anti-de Sitter space applications by considering the universal covering space, in which $t \in \mathbb{R}$.} \, \theta_1, \, \overline{\theta}_1 \in \left[0, \, \pi\right]$, and $\theta_2, \, \phi_1, \, \phi_2, \, \phi_3, \, \overline{\theta}_2, \, \overline{\phi}_1, \, \overline{\phi}_2 \in \left[0, \, 2\pi\right)$. The corresponding line element (for $y^m \equiv \left(t, \, \rho, \, \theta_1, \, \theta_2, \, \phi_1, \, \phi_2, \, \phi_3\right)$, $x^m \equiv \left(\overline{\theta}_1, \, \overline{\theta}_2, \, \overline{\phi}_1, \, \overline{\phi}_2\right)$) is:

\begin{IEEEeqnarray}{rrl}
ds^2 = &G&_{mn}^{AdS}(y) dy^m dy^n + G_{mn}^S(x) dx^m dx^n = \nonumber \\[6pt]
= & 4 \Big[& -\cosh^2\rho \, dt^2 + d\rho^2 + \sinh^2\rho \, \Big(d\theta_1^2 + \cos^2\theta_1 \, d\phi_1^2 + \sin^2\theta_1 \, \big(d\theta_2^2 + \cos^2\theta_2 d\phi_2^2 + \nonumber \\[6pt]
&& + \sin^2\theta_2 d\phi_3^2\big)\Big)\Big] + \Big[d\overline{\theta}_1^2 + \cos^2\overline{\theta}_1 \, d\overline{\phi}_1^2 + \sin^2\overline{\theta}_1 \, \left(d\overline{\theta}_2^2 + \cos^2\overline{\theta}_2 \, d\overline{\phi}_2^2\right)\Big]. \label{LineElement}
\end{IEEEeqnarray} \\
Action \ref{Polyakov1} becomes: \\ [6pt]
\begin{IEEEeqnarray}{ll}
\mathcal{S} = \frac{T_2}{2} \int \Big[& G_{mn}^{AdS}(y) \dot{y}^m \dot{y}^n + G_{mn}^S(x) \dot{x}^m \dot{x}^n - \frac{1}{2} G_{mn}^{AdS}(y) G_{pq}^{AdS}(y) \{y^m, y^p\} \{y^n, y^q\} - \label{PolyakovAction} \\[6pt]
& -\frac{1}{2} G_{mn}^S(x) G_{pq}^S(x) \{x^m, x^p\} \{x^n, x^q\} - G_{mn}^{AdS}(y) G_{pq}^S(x) \{y^m, x^p\} \{y^n, x^q\} \Big] d\tau \, d\sigma \, d\delta, \nonumber
\end{IEEEeqnarray} \\[6pt]
while the constraints that follow from fixing the gauge \ref{MembraneConstraints1}, \ref{MembraneConstraints2} are ($i,j = 1,2$): \\
\begin{IEEEeqnarray}{ll}
\gamma_{00} = - &\det h_{ij} \Rightarrow G_{mn}^{AdS}(y) \dot{y}^m \dot{y}^n + G_{mn}^S(x) \dot{x}^m \dot{x}^n + \frac{1}{2} G_{mn}^{AdS}(y) G_{pq}^{AdS}(y) \{y^m, y^p\} \{y^n, y^q\} + \hspace{.5cm} \nonumber \\[6pt]
& + \frac{1}{2} G_{mn}^S(x) G_{pq}^S(x) \{x^m, x^p\} \{x^n, x^q\} + G_{mn}^{AdS}(y) G_{pq}^S(x) \{y^m, x^p\} \{y^n, x^q\}= 0 \qquad \label{PolyakovConstraints1}
\end{IEEEeqnarray}
\begin{IEEEeqnarray}{ll}
\gamma_{0i} = G_{mn}^{AdS}(y) \, \dot{y}^m \partial_i y^n + G_{mn}^{S}(x) &\, \dot{x}^m \partial_i x^n = 0 \Rightarrow \nonumber \\[6pt]
& \Rightarrow \Big\{G_{mn}^{AdS}(y) \, \dot{y}^m, y^n\Big\} + \Big\{G_{mn}^{S}(x) \, \dot{x}^m, x^n\Big\} = 0. \hspace{1cm} \label{PolyakovConstraints2} \hspace{0.7cm}
\end{IEEEeqnarray}
\newpage
Action \ref{PolyakovAction}, and its constraints \ref{PolyakovConstraints1}--\ref{PolyakovConstraints2}, are invariant under the global isometry $SO(6,2) \times SO(5)$ of AdS$_7 \times S^4$. The following 28+10 Noether charges are conserved on-shell: \\[6pt]
\begin{IEEEeqnarray}{ll}
S^{\mu\nu} = T_2 \int_0^{2\pi}\left(Y^\mu \dot{Y}^\nu - Y^\nu \dot{Y}^\mu\right) \, d\sigma d\delta, \quad & \quad \mu \,,\nu = 0,1,\ldots,7 \\[6pt]
J^{ij} = T_2 \int_0^{2\pi}\left(X^i \dot{X}^j - X^j \dot{X}^i\right) \, d\sigma d\delta, \quad & \quad i,j = 1,2,\ldots,5.
\end{IEEEeqnarray} \\[6pt]
The charges that correspond to the cyclic coordinates of the action \ref{PolyakovAction}, $t, \, \phi_1, \, \phi_2, \, \phi_3, \, \overline{\phi}_1, \, \overline{\phi}_2$, are simpler in form and can be directly read off from \ref{LineElement}--\ref{PolyakovAction}: \\
\begin{IEEEeqnarray}{l}
E = \left|\frac{\partial L}{\partial \dot{t}}\right| = 4\,T_2 \int_0^{2\pi} \dot{t} \cosh^2\rho \, d\sigma d\delta = S^{07} \label{MembraneCyclicCharge1} \\[6pt]
S_1 = \frac{\partial L}{\partial \dot{\phi}_1} = 4\,T_2 \int_0^{2\pi} \dot{\phi}_1 \sinh^2\rho \, \cos^2\theta_1 \, d\sigma d\delta = S^{12} \label{MembraneCyclicCharge2} \\[6pt]
S_2 = \frac{\partial L}{\partial \dot{\phi}_2} = 4\,T_2 \int_0^{2\pi} \dot{\phi}_2 \sinh^2\rho \, \sin^2\theta_1 \cos^2\theta_2 \, d\sigma d\delta = S^{34}\label{MembraneCyclicCharge3} \\[6pt]
S_3 = \frac{\partial L}{\partial \dot{\phi}_3} = 4\,T_2 \int_0^{2\pi} \dot{\phi}_3 \sinh^2\rho \sin^2\theta_1 \sin^2\theta_2 \, d\sigma d\delta = S^{56} \label{MembraneCyclicCharge4} \\[18pt]
J_1 = \frac{\partial L}{\partial \dot{\overline{\phi}}_1} = 4\,T_2 \int_0^{2\pi} \dot{\overline{\phi}}_1 \cos^2\overline{\theta}_1 \, d\sigma d\delta = J^{12} \label{MembraneCyclicCharge5} \\[6pt]
J_2 = \frac{\partial L}{\partial \dot{\overline{\phi}}_2} = 4\,T_2 \int_0^{2\pi} \dot{\overline{\phi}}_2 \sin^2\overline{\theta}_1 \cos^2\overline{\theta}_2 \, d\sigma d\delta = J^{34}, \label{MembraneCyclicCharge6}
\end{IEEEeqnarray} \\[6pt]
where L stands for the Lagrangian of the system that is defined as $\mathcal{S} = \int L \, d\tau$. \\
\section[Spinning Membranes and Spinning Strings.]{Spinning Membranes and Spinning Strings.\label{Strings&Membranes}}

We shall now show that the folded closed string of \cite{GubserKlebanovPolyakov02}, rotating in $AdS_3 \subset AdS_5$, has the same action and equations of motion as a specific membrane soliton that spins in $AdS_3 \subset AdS_7 \times S^4$. This result will later be generalized to any string soliton that lives in pure\footnote{I.e. a soliton with no components in $S^5$ whatsoever.} $AdS_5$, for which an equivalent $AdS_7 \times S^4$ membrane soliton will be found. Let us start from the following ansatz for a membrane that rotates in $AdS_3 \times S^1 \subset AdS_7 \times S^4$: \\
\begin{equation}
\begin{array}{ll} \Big\{t = \kappa \tau, \, \rho = \rho(\sigma), \, \phi_1 = \kappa \omega \tau, \, \phi_2 = \phi_3 = \theta_1 = \theta_2 &= 0\Big\} \times \\[12pt] & \times \Big\{\overline{\phi}_1 = \delta , \, \overline{\theta}_1 = \overline{\theta}_2 = \overline{\phi}_2 = 0\Big\}. \end{array} \label{I} \\[12pt]
\end{equation}
It reads, in embedding coordinates ($R = 1$, $\ell = 2$), \\
\begin{IEEEeqnarray}{lll}
Y_0 = 2 \cosh\rho(\sigma) \cos \kappa\tau &, \quad Y_3 = Y_4 = Y_5 = Y_6 = 0 \,, & \qquad X_1 = \cos \delta \nonumber \\[6pt]
Y_1 = 2 \sinh\rho(\sigma) \cos \kappa\omega \tau && \qquad X_2 = \sin \delta \\[6pt]
Y_2 = 2 \sinh\rho(\sigma) \sin \kappa\omega \tau && \qquad X_3 = X_4 = X_5 = 0 \nonumber \\[6pt]
Y_7 = 2 \cosh\rho(\sigma) \sin \kappa\tau. \nonumber
\end{IEEEeqnarray} \\
The Polyakov action \ref{PolyakovAction} and the constraint equation \ref{PolyakovConstraints1} become:\footnote{In $D = 11$ dimensions, a simple relation between the Planck length $\ell_{11}$, the string coupling constant $g_s$ and the string fundamental length can be deduced by dimensionally reducing 11-dimensional supergravity to 10 dimensions,
\begin{equation}
g_s = \left(\frac{R_c}{l_{11}}\right)^{3/2} \,, \qquad \ell_s^2 = \frac{l_{11}^3}{R_c} \quad \longrightarrow \quad g_s = \left(\frac{\ell_{11}}{\ell_s}\right)^3 \nonumber
\end{equation} ($R_c$ being the compactification radius) so that the membrane tension in 11 dimensions becomes, $T_2 = \left[(2\pi)^2 g_s \ell_s^3\right]^{-1}$ \cite{Kiritsis07}.}\\
\begin{IEEEeqnarray}{ll}
\mathcal{S} & = 2 \, T_2 \int \left(-\dot{t}^2 \cosh^2\rho + \dot{\phi}_1^2 \, \sinh^2\rho \, \cos^2\theta_1 - \cos^2\overline{\theta}_1 \, \rho'^2 \, \overline{\phi}_1'^2 \, \left\{\sigma, \delta\right\}^2\right) d\tau d\sigma d\delta = \label{ActionIa} \quad \\[6pt]
& = \frac{2 \, T_1}{\ell_s g_s} \int \left(- \kappa^2 \cosh^2\rho + \kappa^2 \omega^2 \, \sinh^2\rho - \rho'\,^2\right) d\tau d\sigma \label{ActionIb}
\end{IEEEeqnarray}
\begin{IEEEeqnarray}{ccc}
\rho'\,^2 - \kappa^2\left(\cosh^2\rho - \omega^2 \sinh^2\rho\right) = 0 & \qquad & (\text{constraint}). \label{ConstraintI}
\end{IEEEeqnarray} \\
Action \ref{ActionIb} and constraint \ref{ConstraintI} are identical to the on-shell string Polyakov action (written in the conformal gauge) and the Virasoro constraint of the AdS$_3$ folded closed string configuration of \cite{GubserKlebanovPolyakov02}. Were it not for a factor of $\cos^2\overline{\theta}_1 \, \overline{\phi}_1'^2$, the off-shell action \ref{ActionIa} would also coincide with the corresponding off-shell stringy action. However---in action \ref{ActionIa}---it is only $\rho$ that has a nonzero equation of motion and that equation of motion is identical to the stringy one, \ref{GKPEquationI}:
\begin{IEEEeqnarray}{l}
\rho'' + \kappa^2 \left(\omega^2 - 1\right) \sinh\rho \cosh\rho = 0.
\end{IEEEeqnarray} \\
The conserved charges of the membrane action \ref{ActionIa} are also identical to the ones obtained for strings, \ref{GKPIEnergy2}--\ref{GKPISpin2} (for $\omega^2 > 1$): \\
\begin{IEEEeqnarray}{l}
E(\omega) = \frac{16\,T_1}{g_s \ell_s} \cdot \frac{\omega}{\omega^2 - 1} \, \mathbb{E} \left(\frac{1}{\omega^2}\right) \\[6pt]
S(\omega) = \frac{16\,T_1}{g_s \ell_s} \cdot \left(\frac{\omega^2}{\omega^2 - 1} \, \mathbb{E} \left(\frac{1}{\omega^2}\right) - \mathbb{K} \left(\frac{1}{\omega^2}\right)\right) = S_1.
\end{IEEEeqnarray} \\[6pt]
Therefore the two systems are dynamically equivalent. Another (string) solution of \cite{GubserKlebanovPolyakov02} consists of a closed string that oscillates around the center of $AdS_5$. It can also be written in terms of a pulsating, AdS$_7 \times S^4$ membrane as follows: \\
\begin{equation}
\begin{array}{ll} \Big\{t = t\left(\tau\right), \, \rho = \rho\left(\tau\right), \, \theta_1 = \frac{\pi}{2}, \, \theta_2 = \sigma, \, \phi_1 = &\phi_2 = \phi_3 = 0\Big\} \times \\[12pt] &\times \Big\{\overline{\phi}_1 = \delta , \, \overline{\theta}_1 = \overline{\theta}_2 = \overline{\phi}_2 = 0\Big\}. \end{array} \label{III} \\[12pt]
\end{equation}

In embedding coordinates, the ansatz reads: \\
\begin{IEEEeqnarray}{lll}
Y_0 = 2 \cosh\rho(\tau) \cos t\left(\tau\right) &, \quad Y_1 = Y_2 = Y_4 = Y_6 = 0 \,, & \qquad X_1 = \cos \delta \nonumber \\[6pt]
Y_3 = 2 \sinh\rho(\tau) \cos \sigma && \qquad X_2 = \sin \delta \\[6pt]
Y_5 = 2 \sinh\rho(\tau) \sin \sigma && \qquad X_3 = X_4 = X_5 = 0 \nonumber \\[6pt]
Y_7 = 2 \cosh\rho(\tau) \sin t\left(\tau\right) \nonumber
\end{IEEEeqnarray} \\[6pt]
and has the following membrane/string Polyakov action and constraint equation: \\
\begin{IEEEeqnarray}{ll}
\mathcal{S} & = 2 \, T_2 \int \left(-\dot{t}^2 \cosh^2\rho + \dot{\rho}^2 - \sinh^2\rho \, \sin^2\theta_1 \, \cos^2\overline{\theta}_1 \; \theta_2'^2 \, \overline{\phi}_1'^2 \, \left\{\sigma, \delta\right\}^2\right) d\tau d\sigma d\delta = \quad \\[6pt]
& = \frac{2 \, T_1}{\ell_s g_s} \int \left(-\dot{t}^2 \cosh^2\rho + \dot{\rho}^2 - \sinh^2\rho\right) d\tau d\sigma \label{ActionIII}
\end{IEEEeqnarray}
\begin{IEEEeqnarray}{ccc}
\dot{\rho}^2 - \dot{t}^2 \cosh^2\rho + \sinh^2\rho = 0 & \qquad & (\text{constraint}).
\end{IEEEeqnarray} \\
The same comments that were made in the previous case can be repeated here as well. Our stringy membrane is dynamically equivalent to the pulsating string of \cite{GubserKlebanovPolyakov02} with identical equations of motion, \ref{t-equation}, \ref{rho-equation} (with $w = 1$):
\begin{IEEEeqnarray}{l}
\ddot{t} \cosh^2\rho + 2 \, \dot{t} \, \dot{\rho} \, \cosh\rho \sinh\rho = 0 \\[6pt]
\ddot{\rho} + \sinh\rho \cosh\rho \left(\dot{t}^2 + 1\right) = 0.
\end{IEEEeqnarray} \\
Now, all of the previous results can be generalized to any\footnote{A word of caution is due here. Not all ans\"{a}tze are consistent with the conformal gauge. The statements herein presented concern string solitons that are compatible with the choice of the conformal gauge in Polyakov action. It would be interesting to be able to generalize them to the case of the Nambu-Goto action as well, i.e. independently of the gauge choice.} string soliton that rotates in $AdS_5$ and has no rotating counterpart in $S^5$ (dubbed "pure" solitons for convenience). We thus prove: \\

%
\begin{proposition}
Every pure $AdS_5$ string soliton has an equivalent $AdS_7 \times S^4$ membrane soliton (and not vice versa). \label{Proposition1} \\
\end{proposition}

\underline{Proof:} Start with \ref{PolyakovAction} and \ref{PolyakovConstraints1}--\ref{PolyakovConstraints2}, the membrane Polyakov action  in $AdS_7 \times S^4$ (in the gauge, $\gamma_{00} = - \det h_{ij}$, $\gamma_{0i} = 0$, $\gamma_{ij} = h_{ij}$) and its constraint equations: \\
\begin{IEEEeqnarray}{ll}
\mathcal{S}_2 = \frac{T_2}{2} \int \Big[& G_{mn}^{AdS}(y) \dot{y}^m \dot{y}^n + G_{mn}^S(x) \dot{x}^m \dot{x}^n - \frac{1}{2} G_{mn}^{AdS}(y) G_{pq}^{AdS}(y) \{y^m, y^p\} \{y^n, y^q\} - \\
& -\frac{1}{2} G_{mn}^S(x) G_{pq}^S(x) \{x^m, x^p\} \{x^n, x^q\} - G_{mn}^{AdS}(y) G_{pq}^S(x) \{y^m, x^p\} \{y^n, x^q\} \Big] d\tau \, d\sigma \, d\delta \nonumber
\end{IEEEeqnarray}
\begin{IEEEeqnarray}{l}
G_{mn}^{AdS}(y) \dot{y}^m \dot{y}^n + G_{mn}^S(x) \dot{x}^m \dot{x}^n + \frac{1}{2} G_{mn}^{AdS}(y) G_{pq}^{AdS}(y) \{y^m, y^p\} \{y^n, y^q\} \nonumber + \\[6pt]
+ \frac{1}{2} G_{mn}^S(x) G_{pq}^S(x) \{x^m, x^p\} \{x^n, x^q\} + G_{mn}^{AdS}(y) G_{pq}^S(x) \{y^m, x^p\} \{y^n, x^q\}= 0
\end{IEEEeqnarray}
\begin{IEEEeqnarray}{c}
G_{mn}^{AdS}(y) \, \dot{y}^m \partial_i y^n + G_{mn}^{S}(x) \, \dot{x}^m \partial_i x^n = \Big\{G_{mn}^{AdS}(y) \, \dot{y}^m, y^n\Big\} + \Big\{G_{mn}^{S}(x) \, \dot{x}^m, x^n\Big\} = 0, \qquad
\end{IEEEeqnarray} \\[6pt]
where $y^m \equiv \left(t, \, \rho, \, \theta_1, \, \theta_2, \, \phi_1, \, \phi_2, \, \phi_3\right)$ and $x^m \equiv \left(\overline{\theta}_1, \, \overline{\theta}_2, \, \overline{\phi}_1, \, \overline{\phi}_2\right)$. $G_{mn}\left(y,\,x\right)$ are the components of the metric \ref{LineElement}. Taking $\sigma$ as the string world-sheet coordinate,
\begin{IEEEeqnarray}{c}
y^m = y^m \left(\tau \,, \sigma\right) \quad \& \quad x^m = x^m \left(\tau \,, \delta\right),
\end{IEEEeqnarray}
immediately gives: \\
\begin{IEEEeqnarray}{c}
\mathcal{S}_2 = \frac{T_2}{2} \int \Big[G_{mn}^{AdS}(y) \dot{y}^m \dot{y}^n + G_{mn}^S(x) \dot{x}^m \dot{x}^n - G_{mn}^{AdS}(y) G_{pq}^S(x) y'\,^m y'\,^n x'\,^p x'\,^q \Big] d\tau \, d\sigma \, d\delta \hspace{1cm} \\[12pt]
G_{mn}^{AdS}(y) \dot{y}^m \dot{y}^n + G_{mn}^S(x) \dot{x}^m \dot{x}^n + G_{mn}^{AdS}(y) G_{pq}^S(x) y'\,^m y'\,^n x'\,^p x'\,^q = 0 \\[12pt]
G_{mn}^{AdS}(y) \, \dot{y}^m y'\,^n = G_{mn}^{S}(x) \, \dot{x}^m x'\,^n = 0.
\end{IEEEeqnarray}
\newpage
Choosing $x^3 = \overline{\phi}_1 = \delta$ for the coordinate of $S^4$ with metric component $G_{33}^S = \cos^2\overline{\theta}_1$, \\[6pt]
\begin{IEEEeqnarray}{ll}
\mathcal{S}_2 = \frac{T_2}{2} \int \Big[G_{mn}^{AdS}(y) \Big(\dot{y}^m \dot{y}^n - \cos^2\overline{\theta}_1 &\, {\overline{\phi}_1}'^2 \, y'\,^m y'\,^n\Big) + G_{mn \neq 3}^S(x) \dot{x}^m \dot{x}^n - \nonumber \\[6pt]
& - G_{mn}^{AdS}(y) G_{pq \neq 3}^S(x) y'\,^m y'\,^n x'\,^p x'\,^q \Big] d\tau \, d\sigma \, d\delta \qquad
\end{IEEEeqnarray}
\begin{IEEEeqnarray}{rl}
G_{mn}^{AdS}(y) \left(\dot{y}^m \dot{y}^n + \, \cos^2\overline{\theta}_1 \, y'\,^m y'\,^n\right) + G_{mn \neq 3}^S&(x) \dot{x}^m \dot{x}^n + \nonumber \\[6pt]
& + G_{mn}^{AdS}(y) G_{pq \neq 3}^S(x) y'\,^m y'\,^n x'\,^p x'\,^q = 0 \hspace{1cm} \\[12pt]
G_{mn}^{AdS}(y) \, \dot{y}^m y'\,^n = &G_{mn \neq 3}^{S}(x) \, \dot{x}^m x'\,^n = 0.
\end{IEEEeqnarray} \\[6pt]
The result \ref{Proposition1} follows upon setting $x^{m \neq 3} = 0$, $y^{m > 5} = 0$, and performing the $\delta$-integration: \\[3pt]
\begin{IEEEeqnarray}{ll}
\mathcal{S}_2 &= \frac{T_2}{2} \int \, G_{mn \leq 5}^{AdS}(y^{p \leq 5}) \left(\dot{y}^m \dot{y}^n - \cos^2\overline{\theta}_1 \, {\overline{\phi}_1}'^2 \, y'\,^m y'\,^n\right) d\tau \, d\sigma = \label{MembraneAction}\\[6pt]
&= \frac{T_1}{2 g_s \ell_s} \int \, G_{mn \leq 5}^{AdS}(y^{p \leq 5}) \left(\dot{y}^m \dot{y}^n - y'\,^m y'\,^n\right) d\tau \, d\sigma = \frac{\mathcal{S}_1}{g_s \ell_s}
\end{IEEEeqnarray}

\begin{IEEEeqnarray}{c}
G_{mn \leq 5}^{AdS}(y^{p \leq 5}) \left(\dot{y}^m \dot{y}^n + y'\,^m y'\,^n\right) = G_{mn \leq 5}^{AdS}(y^{p \leq 5}) \, \dot{y}^m y'\,^n = 0,
\end{IEEEeqnarray} \\[3pt]
i.e. a pure $AdS_5$ string soliton. For comparison, we juxtapose the corresponding $AdS_5 \times S^5$ string Polyakov action in the conformal gauge ($\gamma_{ab} = \eta_{ab}$) and the corresponding Virasoro constraints: \\[6pt]
\begin{IEEEeqnarray}{c}
\mathcal{S}_1 = \frac{T_1}{2} \int \Big[G_{mn}^{AdS}(y) \left(\dot{y}^m \dot{y}^n - y'\,^m y'\,^n\right) + G_{mn}^S(x)\left(\dot{x}^m \dot{x}^n - x'\,^m x'\,^n\right)\Big] d\tau \, d\sigma \label{StringAction}
\end{IEEEeqnarray}

\begin{IEEEeqnarray}{l}
T_{00} = T_{11} = \frac{1}{2} \Big[G_{mn}^{AdS}(y) \left(\dot{y}^m \dot{y}^n + y'\,^m y'\,^n\right) + G_{mn}^S(x)\left(\dot{x}^m \dot{x}^n + x'\,^m x'\,^n\right)\Big] = 0 \\ [24pt]
T_{01} = T_{10} = G_{mn}^{AdS}(y) \, \dot{y}^m y'\,^n + G_{mn}^S(x) \, \dot{x}^m x'\,^n = 0.
\end{IEEEeqnarray} \\[6pt]
The equations of motion that correspond to $\overline{\theta}_1$ and $\overline{\phi}_1$ in \ref{MembraneAction} are trivially satisfied and the remaining equations of motion of \ref{MembraneAction} will be identical to the ones that are obtained by varying the string action \ref{StringAction}. Thus the two systems are dynamically equivalent. \\[6pt]
The not vice versa part in \ref{Proposition1} follows from the fact that we may construct many inequivalent membrane actions with dependence on both $\sigma$ and $\delta$. $\Box$ \\[6pt]
\subsection[Stringy Membranes in $AdS_4 \times S^7$.]{Stringy Membranes in $AdS_4 \times S^7$. \label{AdS4xS7}\\}

Going over to the $AdS_4 \times S^7$ case, construction \ref{Proposition1} has to be modified in the following way. Assuming complete dependence of the string's spacetime coordinates on the world-sheet coordinates $\left\{\tau \,, \sigma\right\}$, \\
\begin{equation}
y^m = \left(t = t\left(\tau\,,\sigma\right), \, \rho = \rho\left(\tau\,,\sigma\right), \, \theta = \theta\left(\tau\,,\sigma\right), \, \phi_1 = \phi_1\left(\tau\,,\sigma\right), \, \phi_2 = \phi_2\left(\tau\,,\sigma\right)\right), \label{Coordinates} \\[12pt]
\end{equation}

it is only a subset of all possible $AdS_5$ string solitons that can be obtained from an appropriate membrane ansatz on $AdS_4 \times S^7$ --- namely all string solitons that live in $AdS_4 \subset AdS_5$. For example, both stringy anti-de Sitter solitons encountered in this paper (\ref{I}, \ref{III}) are of this genre, living in $AdS_3 \subset AdS_4 \subset AdS_5$. Thus, they can be reproduced by an $AdS_4 \times S^7$ membrane: \\
%
\begin{proposition}
Every pure string soliton of $AdS_4 \subset AdS_5$\footnote{AdS$_4 \subset$ AdS$_5$ means that one of the two azimuthal angles of $S^3$ of AdS$_5$ is set to zero.} has an equivalent $AdS_4 \times S^7$ membrane soliton (and not vice versa). \label{Proposition3} \\
\end{proposition}
Dropping the condition of full dependence on the world-sheet coordinates \ref{Coordinates}, it should be possible to apply this method and find, (i) $AdS_{4,7} \times S^{7,4}$ membrane equivalents to special string configurations that live in $AdS_5 \times S^5$ and (ii) $AdS_4 \times S^7$ membranes that are equivalent to strings that live in $AdS_5$. \\[6pt]
\subsection[Stringy Membranes in $AdS_4 \times S^7/\mathbb{Z}_k$.]{Stringy Membranes in $AdS_4 \times S^7/\mathbb{Z}_k$.\label{AdS4xS7/Zk}\\}

We can also consider stringy membranes in more general backgrounds, such as $AdS_4 \times S^7/\mathbb{Z}_k$. For $k = 1$, this is just $AdS_4 \times S^7$. On the other hand, $AdS_4 \times S^7/\mathbb{Z}_k$ geometries provide the gravitational backgrounds of the ABJM correspondence \cite{ABJM08}:

\begin{center}
\rule{0pt}{4ex} \Big\{$\mathcal{N} = 6$,  $U\left(N\right)_k \times U\left(N\right)_{-k}$, Super C-S Theory\footnote{Super Chern-Simons theory.}\Big\} $\xrightarrow{N \rightarrow\infty}$ \Big\{M-Theory on $AdS_4 \times S^7/\mathbb{Z}_k$\Big\}. \\[24pt]
\end{center}
For $k = 1$, the dual gauge theory reduces to $\mathcal{N} = 8$ SCFT. In the case of the $SU\left(2\right) \times SU\left(2\right)$ gauge group, it becomes the $\mathcal{N} = 8$ Bagger-Lambert-Gustavsson (BLG) theory, \cite{BaggerLambert07, Gustavsson07}. Now the question has been posed, whether a logarithmic type behavior for the anomalous dimensions of either theory's states is possible within this correspondence as well. Based on what has been said above, the answer is affirmative from the point of view of membranes. To see this, consider the metric of $AdS_4 \times S^7/\mathbb{Z}_k$ \cite{BergshoeffDuffPopeSezgin89}: \\
\begin{IEEEeqnarray}{ll}
ds^2 &= G_{mn}^{AdS}(y) dy^m dy^n + G_{mn}^{S/\mathbb{Z}}(x) dx^m dx^n = \nonumber \\[6pt]
&= \ell^2 \left(-\cosh^2\rho \, dt^2 + d\rho^2 + \sinh^2\rho \cdot d\Omega_2^2\right) + R^2 d\overline{\Omega}_{7/\mathbb{Z}_k}^2 \hspace{3.3cm}
\end{IEEEeqnarray}
\begin{IEEEeqnarray}{ll}
d\overline{\Omega}_{7/\mathbb{Z}_k}^2 = &\left(\frac{d\overline{y}}{k} + \widetilde{A}\right)^2 + ds_{\text{CP}^3}^2 \,, \\[6pt]
&\widetilde{A} \equiv \frac{1}{2} \left(\cos^2 \overline{\xi} - \sin^2 \overline{\xi}\right) d\overline{\psi} + \frac{1}{2} \cos^2\overline{\xi} \, \cos \overline{\theta}_1 \, d\overline{\phi}_1 + \frac{1}{2}\sin^2\overline{\xi} \, \cos\overline{\theta}_2 \, d\overline{\phi}_2 \nonumber
\end{IEEEeqnarray}
\begin{IEEEeqnarray}{ll}
ds_{\text{CP}^3}^2 = d\overline{\xi}^2 &+ \cos^2\overline{\xi} \, \sin^2\overline{\xi} \left(d\overline{\psi} + \frac{1}{2} \cos\overline{\theta}_1 \, d\overline{\phi}_1 - \frac{1}{2} \cos\overline{\theta}_2 \, d\overline{\phi}_2\right)^2 + \nonumber \\[6pt]
&+ \frac{1}{4} \cos^2\overline{\xi} \Big(d\overline{\theta}_1^2 + \sin^2 \overline{\theta}_1 \, d\overline{\phi}_1^2\Big) + \frac{1}{4}\sin^2\overline{\xi} \left(d\overline{\theta}_2^2 + \sin^2\overline{\theta}_2 \, d\overline{\phi}_2^2\right).
\end{IEEEeqnarray} \\[6pt]
It is easy to obtain solutions \ref{I} and \ref{III} from this metric. All that is needed is to supplement the AdS ans\"{a}tze with $\overline{y} = k \delta$ ($R = 1$, $\ell = 1 /2$) and set the six remaining angles of $S^7$ equal to zero. In fact, one could formulate the following proposition: \\
%
\begin{proposition}
Every pure string soliton of $AdS_4 \subset AdS_5$ has an equivalent $AdS_4 \times S^7/\mathbb{Z}_k$ membrane soliton (and not vice versa). \label{Proposition4} \\
\end{proposition}
Of course, more general statements than \ref{Proposition4} exist, since type IIA string theory action on $AdS_4 \times \text{CP}^3$ is obtainable from the supermembrane action on $AdS_4 \times S^7$ by double dimensional reduction \cite{ArutyunovFrolov08, GomisSorokinWulff08, Uvarov09}. \\[6pt]
This concludes our presentation of anti-de Sitter space stringy membranes. In the following section we shall examine their stability properties. \\
\section[Membrane Fluctuations.]{Membrane Fluctuations.\label{Fluctuations}}

Are stringy membranes stable? Intuitively, one would expect that the $\delta$-component of a stringy membrane, that is wound around a great circle of $S^{4/7}$, would be unstable towards a lower energy configuration that is obtained by its collapsing to a point on either pole. This would indeed be the case for the simplest string extending along a great circle of a sphere and has no other dynamical parts \cite{FrolovTseytlin03a}. Besides, since stringy membranes share a common Lagrangian and equations of motion with their equivalent strings, they are expected to inherit many of their stabilities/instabilities. Now, unstable strings may be stabilized in a multitude of ways, e.g. by adding more angular momenta \cite{FrolovTseytlin03a, FrolovTseytlin03b}, stable AdS components \cite{TirziuTseytlin09b, ArutyunovRussoTseytlin03}, pulsation \cite{KhanLarsen05}, by orientifold projections \cite{Stefanski03}, or even flux terms \cite{MaldacenaOoguri00, BachasDouglasSchweigert00}\nocite{BachasPetropoulos00}. Surprisingly enough, even those stringy configurations that are known to possess unstable modes, have been studied and have been proven very useful in the context of AdS/CFT \cite{FrolovTseytlin03a, ArutyunovRussoTseytlin03}, as their instabilities are sometimes unseen in the dual gauge sector \cite{FrolovTseytlin03c, BeisertMinahanStaudacherZarembo03}. One possible explanation for this state of affairs is that these solutions can be easily extended to more stable configurations, while preserving their wanted dual gauge theory properties. Generalized, rigorous results (even numerical) concerning stability are however missing at the moment, mainly due to the difficulties that the corresponding analysis presents \cite{FrolovTirziuTseytlin06}. \\[6pt]
Stringy membranes are on the other hand membranes, not strings. We believe that this property may sometimes enhance the stability of the resulting system. For example, a single membrane component that is wound around a sphere has zero surface tension and is thus expected to be stable, in contrast to the similarly wound string that we saw above. Since we are actually proposing a model that attempts to reproduce the behavior of classical strings in AdS$_5$, it would be interesting to be able to make concrete statements about its advantages/disadvantages in the domain of stability. Membrane fluctuations in various backgrounds have been studied in \cite{Lousto95, LarsenLousto96a, LarsenLousto96b, HarmarkSavvidy00, SavvidySavvidy00, AxenidesFloratosPerivolaropoulos00, AxenidesFloratosPerivolaropoulos01, Savvidy01, AxenidesFloratosPerivolaropoulos02}. \\[6pt]
Interestingly, we shall find that our systems are governed by the Lam\'{e} equation. Lam\'{e} equations arise when one separates variables in Laplace's equation using an ellipsoidal coordinate system \cite{Lame37}. They belong to the class of the so-called quasi-exactly solvable (QES) systems \cite{Turbiner88, Ushveridze94}, because their solutions may be determined algebraically in some cases \cite{AlhassidGurseyIachello83, LiKusnezov99, LiKusnezovIachello99, FinkelGonzalez-LopezRodriguez99, Maier03}. Owing to the fact that their stabilities and instabilities are organized in bands and gaps, Lam\'{e} systems enjoy a wide range of physical applications: (a) they provide an alternative to the Kronig-Penney model for the motion of electrons in one-dimensional crystals \cite{Sutherland73, AlhassidGurseyIachello83}; (b) they govern explosive particle production (preheating) due to parametric resonance in post-inflationary cosmology \cite{KofmanLindeStarobinsky94, BoyanovskydeVegaHolmanSalgado96, GreeneKofmanLindeStarobinsky97}; (c) they arise in the study of sphaleron fluctuations in the $\phi^4$ \cite{MantonSamols88, LiangMuller-KirstenTchrakian92} and 1+1 dimensional abelian Higgs model \cite{BrihayeGillerKosinskiKunz92, BraibantBrihaye93}; (d) they are closely related to the spectral curve of $SU(2)$ BPS monopoles \cite{Ward87, Sutcliffe96}; (e) they come up in many occasions in supersymmetric quantum mechanics \cite{DunneFeinberg97, DunneMannix97, KhareSukhatme99, CorreaPlyushchay07}, etc. \cite{FloratosNicolis95, BakasBrandhuberSfetsos99, BakasBrandhuberSfetsos00, Dunne02, DunneShifman02}. They have also appeared in string fluctuations in anti-de Sitter space \cite{KhanLarsen05, Beccariaetal10a, Beccariaetal10b}. The examination of stringy membranes in the present work, is suggestive of a much richer Lam\'{e} band/gap structure for their fluctuations. We will have more to say about the stabilities and instabilities of stringy membranes at the end of this section. \\[6pt]
Our analysis will be carried out in the embedding coordinates of AdS$_{p + 2} \times S^q$, for which, \\
\begin{IEEEeqnarray}{c}
ds^2 = \eta_{\mu\nu} dY^\mu dY^\nu + \delta_{ij} dX^i dX^j = - dY_0^2 + \sum_{i = 1}^{p + 1} dY_i^2 - dY_{p + 2}^2 + \sum_{i = 1}^{q + 1} dX_i^2 \label{EmbeddingCoordinates1} \\[6pt]
- \eta_{\mu\nu} Y^\mu Y^\nu = Y_0^2 - \sum_{i = 1}^{p + 1} Y_i^2 + Y_{p + 2}^2 = \ell^2 \quad , \quad \delta_{ij} X^i X^j = \sum_{i = 1}^{q + 1} X_i^2 = R^2, \label{EmbeddingCoordinates2}
\end{IEEEeqnarray} \\[6pt]
where $\eta_{\mu\nu} = \left(-,+,+,\ldots,+,-\right)$, $\delta_{ij} = \left(+,+,\ldots,+\right)$, $\mu,\nu = 0,1,\ldots,p + 2$ and $i,j = 1,2,\ldots,q + 1$. Including the constraints \ref{EmbeddingCoordinates2} with the aid of two Lagrange multipliers $\Lambda$, $\widetilde{\Lambda}$, the gauge-fixed action \ref{Polyakov1} becomes: \\
\begin{IEEEeqnarray}{ll}
\mathcal{S} = \frac{T_2}{2} \int d^3\sigma \Bigg[&\dot{Y}^\mu \dot{Y}_\mu + \dot{X}^i \dot{X}^i - \frac{1}{2} \{Y^\mu, Y^\nu\} \{Y_\mu, Y_\nu\}  - \frac{1}{2} \{X^i, X^j\} \{X^i, X^j\} - \nonumber \\[6pt]
& - \{Y^\mu, X^i\} \{Y_\mu, X^i\} + \Lambda \left(Y^\mu Y_\mu + \ell^2\right) + \widetilde{\Lambda} \left(X^i X^i - R^2\right)\Bigg].
\end{IEEEeqnarray}\\
It gives rise to the following equations of motion: \\
\begin{IEEEeqnarray}{c}
\ddot{Y}^\mu = \left\{\left\{Y^\mu,Y^\nu\right\},Y_\nu\right\} + \left\{\left\{Y^\mu,X^i\right\},X^i\right\} + \Lambda \, Y^\mu  \label{EmbeddingEoM1} \\[18pt]
\ddot{X}^i = \left\{\left\{X^i,X^j\right\},X^j\right\} + \left\{\left\{X^i,Y^\mu\right\},Y_\mu\right\} + \widetilde{\Lambda} \, X^i,  \label{EmbeddingEoM2}
\end{IEEEeqnarray} \\
the AdS$_{p +2} \times S^q$ constraints,
\begin{IEEEeqnarray}{c}
Y^\mu Y_\mu = - \ell^2 \quad , \quad X^i X^i = R^2  \label{EmbeddingEoM3}
\end{IEEEeqnarray}

and the constraints that follow from gauge-fixing, \\
\begin{IEEEeqnarray}{c}
\dot{Y}^\mu \partial_\sigma Y_\mu + \dot{X}^i \partial_\sigma X^i = \dot{Y}^\mu \partial_\delta Y_\mu + \dot{X}^i \partial_\delta X^i = 0  \label{EmbeddingEoM4} \\[18pt]
\dot{Y}^\mu \dot{Y}_\mu + \dot{X}^i \dot{X}^i + \frac{1}{2} \{Y^\mu, Y^\nu\} \{Y_\mu, Y_\nu\} + \frac{1}{2} \{X^i, X^j\} \{X^i, X^j\} + \{Y^\mu, X^i\} \{Y_\mu, X^i\} = 0.  \qquad \label{EmbeddingEoM5}
\end{IEEEeqnarray} \\
The Hamiltonian is identically equal to zero and it is given by: \\
\begin{IEEEeqnarray}{ll}
H = \frac{T_2}{2} \int d^2\sigma \Bigg[&\dot{Y}^\mu \dot{Y}_\mu + \dot{X}^i \dot{X}^i + \frac{1}{2} \{Y^\mu, Y^\nu\} \{Y_\mu, Y_\nu\} + \frac{1}{2} \{X^i, X^j\} \{X^i, X^j\} + \nonumber \\[6pt]
& + \{Y^\mu, X^i\} \{Y_\mu, X^i\} - \Lambda \left(Y^\mu Y_\mu + \ell^2\right) - \widetilde{\Lambda} \left(X^i X^i - R^2\right)\Bigg] = 0.
\end{IEEEeqnarray} \\
We now consider the following perturbations:\footnote{In this section, due care should be taken in order not to confuse $\delta \equiv \sigma_2$, the world-volume coordinate, with the $\delta$'s that appear in $\delta \mathcal{S}$, $\delta X$, $\delta Y$, $\delta \Lambda$, $\delta \widetilde{\Lambda}$ and denote the fluctuations of $\mathcal{S}$, X, Y, $\Lambda$ and $\widetilde{\Lambda}$.}
\begin{IEEEeqnarray}{c}
Y^\mu = Y^\mu_0 + \delta Y^\mu \quad , \quad X^i = X^i_0 + \delta X^i \quad , \quad \Lambda = \Lambda_0 + \delta \Lambda \quad , \quad \widetilde{\Lambda} = \widetilde{\Lambda}_0 + \delta \widetilde{\Lambda},
\end{IEEEeqnarray}
\newpage
where $\left\{Y_0, X_0, \Lambda_0, \widetilde{\Lambda}_0\right\}$ is a classical solution that satisfies the equations of motion and constraints \ref{EmbeddingEoM1}--\ref{EmbeddingEoM5}. The (quadratic) action for the fluctuations is: \\
\begin{IEEEeqnarray}{ll}
\delta \mathcal{S} = \frac{T_2}{2} \int d^3\sigma \Bigg[& \delta\dot{Y}^\mu \, \delta\dot{Y}_\mu + \delta\dot{X}^i \, \delta\dot{X}^i - \{Y_0^\mu, Y_0^\nu\} \{\delta Y_\mu, \delta Y_\nu\} - \{\delta Y^\mu, Y_0^\nu\} \{\delta Y_\mu, Y_{0\,\nu}\} - \nonumber \\[12pt]
& - \{\delta Y^\mu, Y_0^\nu\} \{Y_{0\,\mu}, \delta Y_\nu\} - \{X_0^i, X_0^j\} \{\delta X^i, \delta X^j\} - \{\delta X^i, X_0^j\} \{\delta X^i, X_0^j\} -  \nonumber \\[18pt]
& - \{\delta X^i, X_0^j\} \{X_0^i, \delta X^j\} - 2\{Y_0^\mu, X_0^i\} \{\delta Y_\mu, \delta X^i\} - \{\delta Y^\mu, X_0^i\} \{\delta Y_\mu, X_0^i\} - \nonumber \\[18pt]
& - 2\{\delta Y^\mu, X_0^i\} \{Y_{0\,\mu}, \delta X^i\} - \{Y_0^\mu, \delta X^i\} \{Y_{0\,\mu}, \delta X^i\} + 2 \, Y_0^\mu \, \delta Y_\mu \, \delta\Lambda + \nonumber \\[12pt]
& + 2 \, X_0^i \, \delta X^i \, \delta\widetilde{\Lambda}\Bigg].
\end{IEEEeqnarray} \\[6pt]
To lowest order, these fluctuations obey the following equations: \\
\begin{IEEEeqnarray}{ll}
\delta\ddot{Y}^\mu = &\left\{\left\{Y_0^\mu,Y_0^\nu\right\},\delta Y_\nu\right\} + \left\{\left\{\delta Y^\mu,Y_0^\nu\right\},Y_{0\,\nu}\right\} + \left\{\left\{Y_0^\mu,\delta Y^\nu\right\},Y_{0\,\nu}\right\} + \left\{\left\{Y_0^\mu,X_0^i\right\},\delta X^i\right\} + \nonumber \\[12pt]
& + \left\{\left\{\delta Y^\mu,X_0^i\right\},X_0^i\right\} + \left\{\left\{Y_0^\mu,\delta X^i\right\},X^i\right\} + \Lambda_0 \delta Y^\mu + Y_0^\mu \, \delta\Lambda \label{FluctuationEquation1}
\end{IEEEeqnarray}

\begin{IEEEeqnarray}{ll}
\delta\ddot{X}^i = &\left\{\left\{X_0^i,X_0^j\right\},\delta X_j\right\} + \left\{\left\{\delta X^i,X_0^j\right\},X_0^j\right\} + \left\{\left\{X_0^i,\delta X^j\right\}, X_0^j\right\} + \left\{\left\{X_0^i,Y_0^\mu\right\},\delta Y_\mu\right\} + \nonumber \\[12pt]
& + \left\{\left\{\delta X^i,Y_0^\mu\right\},Y_{0\,\mu}\right\} + \left\{\left\{X_0^i,\delta Y^\mu\right\},Y_{0\,\mu}\right\} + \widetilde{\Lambda}_0 \delta X^i + X_0^i \, \delta\widetilde{\Lambda} \label{FluctuationEquation2}
\end{IEEEeqnarray} \\
and the following constraints (note that our fluctuations live in tangent space): \\
\begin{IEEEeqnarray}{ll}
Y_0^\mu \, \delta Y_\mu = X_0^i \, \delta X^i = 0 \quad , \quad & \dot{Y}_0^\mu \, \partial_\sigma \delta Y_\mu + \delta\dot{Y}^\mu \, \partial_\sigma Y_{0\,\mu} + \dot{X}_0^i \, \partial_\sigma \delta X^i + \delta\dot{X}^i \, \partial_\sigma X_0^i = 0 \nonumber \\[12pt]
& \dot{Y}_0^\mu \, \partial_\delta \delta Y_\mu + \delta\dot{Y}^\mu \, \partial_\delta Y_{0\,\mu} + \dot{X}_0^i \, \partial_\delta \delta X^i + \delta\dot{X}^i \, \partial_\delta X_0^i = 0 \label{FluctuationConstraint1} \qquad
\end{IEEEeqnarray}

\begin{IEEEeqnarray}{l}
\dot{Y}_0^\mu \, \delta\dot{Y}_\mu + \dot{X}_0^i \, \delta\dot{X}^i + \{Y_0^\mu, Y_0^\nu\} \{\delta Y_\mu, Y_{0\,\nu}\} + \{X_0^i, X_0^j\} \{\delta X^i, X_0^j\} + \{Y_0^\mu, X_0^i\} \{\delta Y_\mu, X_0^i\} + \nonumber \\[12pt]
+ \{Y_0^\mu, X_0^i\} \{Y_{0\,\mu}, \delta X^i\} = 0. \label{FluctuationConstraint2}
\end{IEEEeqnarray} \\[6pt]
In order to pass from the general case of an M2-brane in AdS$_{p + 2} \times S^q$ to the general case of a stringy membrane in AdS$_{p + 2} \times S^q$ (i.e. before considering any particular ansatz) we plug, \\
\begin{IEEEeqnarray}{l}
Y^\mu_0 = Y^\mu_0\left(\tau,\sigma\right) \\[12pt]
X_0^i = \left(\cos\delta\,,\,\sin\delta\,,\,0\,,\,\ldots\,,\,0\right) \; \longrightarrow \; X_0^i X_0^i = 1 \\[6pt]
{X_0^i}' = \left(-\sin\delta\,,\,\cos\delta\,,\,0\,,\,\ldots\,,\,0\right) \; \longrightarrow \; {X_0^i}' {X_0^i}' = 1 \\[6pt]
{X_0^i}'' = - \left(\cos\delta\,,\,\sin\delta\,,\,0\,,\,\ldots\,,\,0\right) = - X_0^i  \; \longrightarrow \; {X^i}'' {X^i}'' = 1,
\end{IEEEeqnarray} \\[6pt]
into the equations of the solutions \ref{EmbeddingEoM1}--\ref{EmbeddingEoM5} and those of the fluctuations \ref{FluctuationEquation1}--\ref{FluctuationConstraint2}, setting also $R = 1$. This leads to the following equations of motion, \\
\begin{IEEEeqnarray}{c}
\ddot{Y}_0^\mu = {Y_0^\mu}'' + \Lambda_0 \, Y_0^\mu \quad , \quad {Y_0^\mu}' Y'_{0\,\mu} = - \dot{Y}_0^\mu \dot{Y}_{0\,\mu} = \widetilde{\Lambda}_0 = - \ell^2 / 2 \, \Lambda_0 \label{ZerothOrderEquation1} \\[12pt]
Y_0^\mu Y_{0\,\mu} = - \ell^2 \quad , \quad \dot{Y}_0^\mu Y_{0\,\mu}' = 0, \label{ZerothOrderEquation2}
\end{IEEEeqnarray}
fluctuation equations, \\
\begin{IEEEeqnarray}{ll}
\delta\ddot{Y}^\mu = &\partial_\sigma^2 \delta Y^\mu + \widetilde{\Lambda}_0 \, \partial_\delta^2 \delta Y^\mu - \left({X_0^i}'' \, \partial_\sigma \delta X^i - {X_0^i}' \, \partial_{\sigma,\delta}^2 \delta X^i + {Y_0^\nu}' \, \partial_\delta^2 \delta Y_\nu\right) {Y_0^\mu}' + \nonumber \\[12pt]
&+ 2 \left({X_0^i}' \, \partial_\delta \delta X^i\right) {Y_0^\mu}'' + \Lambda_0 \, \delta Y^\mu + Y_0^\mu \, \delta\Lambda \label{FirstOrderEquation1} \\[18pt]
\delta\ddot{X}^i = &\partial_\sigma^2 \delta X^i + \widetilde{\Lambda}_0 \, \partial_\delta^2 \delta X^i - \left({X_0^j}' \, \partial_\sigma^2 \delta X^j + {Y_0^\mu}'' \, \partial_\delta \delta Y_\mu - {Y_0^\mu}' \, \partial_{\sigma,\delta}^2 \delta Y_\mu\right){X_0^i}' + \nonumber \\[12pt]
&+ 2 \left({Y_0^\mu}' \, \partial_\sigma \delta Y_\mu\right) {X_0^i}'' + \widetilde{\Lambda}_0 \, \delta X^i + X_0^i \, \delta\widetilde{\Lambda} \label{FirstOrderEquation2}
\end{IEEEeqnarray} \\[6pt]
and constraints: \\
\begin{IEEEeqnarray}{c}
Y_0^\mu \, \delta Y_\mu = X_0^i \, \delta X^i = 0 \; , \; \dot{Y}_0^\mu \, \partial_\sigma \delta Y_\mu + \delta\dot{Y}^\mu \, Y'_{0\,\mu} = \dot{Y}_0^\mu \, \partial_\delta \delta Y_\mu + \delta\dot{X}^i \, {X_0^i}' = 0 \qquad \quad \label{FirstOrderEquation3} \\[12pt]
\dot{Y}_0^\mu \, \delta\dot{Y}_\mu + {Y_0^\mu}' \, \partial_\sigma \delta Y_\mu + \widetilde{\Lambda}_0 \,  \left({X_0^i}' \, \partial_\delta \delta X^i\right) = 0. \label{FirstOrderEquation4}
\end{IEEEeqnarray} \\[6pt]
Note that, although the equations of motion \ref{ZerothOrderEquation1}--\ref{ZerothOrderEquation2} are completely independent of the second world-volume coordinate $\delta$ (they are string equations), the fluctuation equations \ref{FirstOrderEquation1}--\ref{FirstOrderEquation4} depend explicitly on $\delta$, through the $S^4$ coordinates $X^i\left(\delta\right)$ and their derivatives. We could not come up with any coordinate transformation that eliminates this dependence on $\delta$. It seems therefore that the equivalence between stringy membranes and strings cannot be extended beyond leading order. \\[6pt]
In order to facilitate further analysis, we shall only study fluctuations along the directions that are transverse to the membrane, i.e. directions for which $Y_0^\mu = X_0^i = 0$. From equations \ref{FirstOrderEquation1}--\ref{FirstOrderEquation4}, we see that these fluctuations decouple from the ones that take place parallel to the stringy membrane. Having said this, the corresponding equations become: \\
\begin{IEEEeqnarray}{l}
\delta\ddot{Y}^\mu = \partial_\sigma^2 \delta Y^\mu + \widetilde{\Lambda}_0 \, \partial_\delta^2 \delta Y^\mu + \Lambda_0 \, \delta Y^\mu \label{FluctuationEquationsY} \\[12pt]
\delta\ddot{X}^i = \partial_\sigma^2 \delta X^i + \widetilde{\Lambda}_0 \, \partial_\delta^2 \delta X^i + \widetilde{\Lambda}_0 \, \delta X^i. \label{FluctuationEquationsX}
\end{IEEEeqnarray} \\
\subsection[Rotating Stringy Membranes.]{Rotating Stringy Membranes.\\}
\begin{figure}
\centering
\includegraphics[scale=0.4]{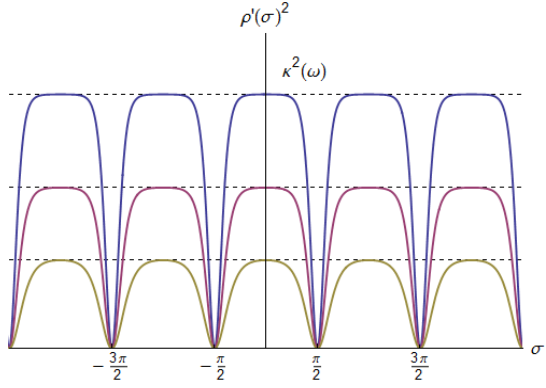}
\caption{Lam\'{e} potential \ref{LamePotentialI} of stringy membrane \ref{I}--\ref{Ia}.} \label{Graph:LamePotentialI}
\end{figure}

To study the fluctuations of spinning stringy membranes we set: \\
\begin{IEEEeqnarray}{l}
\delta Y^\mu = \sum_{r,m} e^{i r \tau + i m \delta} \; \widetilde{y}_{r,m}^{\,\mu}\left(\sigma\right)\,, \qquad \delta X^i = \sum_{r,m} e^{i r \tau + i m \delta} \; \widetilde{x}_{r,m}^{\,i}\left(\sigma\right)\,, \qquad m \in \mathbb{Z}. \label{FluctuationAnsatzI}
\end{IEEEeqnarray} \\[6pt]
In this case, equations \ref{FluctuationEquationsY}--\ref{FluctuationEquationsX} along the transverse directions $Y_0^\mu = X_0^i = 0$, take the following form (omitting, for simplicity, the dependencies of $\widetilde{y}_{r,m}^{\,\mu}\left(\sigma\right)$ and $\widetilde{x}_{r,m}^{\,i}\left(\sigma\right)$ on r, m and $\sigma$): \\
\begin{IEEEeqnarray}{c}
\left(\widetilde{y}^{\,\mu}\right)'' + \left(r^2 - m^2 \widetilde{\Lambda}_0 + \Lambda_0\right) \widetilde{y}^{\,\mu} = 0 \label{FluctuationEquations5} \\[12pt]
\left(\widetilde{x}^{\,i}\right)'' + \left(r^2 - m^2 \widetilde{\Lambda}_0 + \widetilde{\Lambda}_0\right) \widetilde{x}^{\,i} = 0. \label{FluctuationEquations6}
\end{IEEEeqnarray} \\[6pt]
For the AdS$_7 \times S^4$ stringy membranes \ref{I} we have ($\ell = 2$):\footnote{With slight modifications, all the results of this section are also valid in AdS$_4 \times S^7 / \mathbb{Z}^k$. See table \ref{LameParameters}.}\\
\small{\begin{IEEEeqnarray}{c}
Y_0^\mu = 2 \left(\cosh\rho\left(\sigma\right) \cos \kappa\tau \,,\, \sinh\rho\left(\sigma\right) \cos \kappa\omega \tau \,,\, \sinh\rho\left(\sigma\right) \sin \kappa\omega \tau \,,\, 0 \,,\, 0 \,,\, 0 \,,\, 0 \,,\, \cosh\rho\left(\sigma\right) \sin \kappa\tau \right). \qquad \label{Ia}
\end{IEEEeqnarray}} \normalsize \\[6pt]
The equations of motion \ref{ZerothOrderEquation1}--\ref{ZerothOrderEquation2} for this configuration are satisfied for\footnote{Due care should be taken in this subsection, in order to distinguish the elliptic modulus k, from the parameter $\kappa$ of ansatz \ref{Ia} and the complete elliptic integral of the first kind $\mathbb{K}$.}
\begin{IEEEeqnarray}{ll}
\Lambda_0 = - 2 \rho'^2 \quad \& \quad \widetilde{\Lambda}_0 = 4 \rho'^2,
\end{IEEEeqnarray}

where $\rho'\left(\sigma\right)^2$ is the $\sigma$-periodic and even function\footnote{For large enough $\omega$, it turns out that we may approximate $\rho'^2 = \kappa^2 \cdot cd^2\left[\kappa \omega \sigma \Big| 1/ \omega^2\right] \sim \kappa^2 \cos^2 \sigma$.} (displayed for various $\omega$, in figure \ref{Graph:LamePotentialI}), \\
\begin{IEEEeqnarray}{c}
\rho'^2 = \kappa^2 \left(\cosh^2\rho - \omega^2 \sinh^2\rho\right) = \kappa^2 \cdot sn^2\left[\kappa \omega \left(\sigma + \frac{\pi}{2}\right) \Big| \frac{1}{\omega^2}\right] \label{LamePotentialI}\\[6pt]
 \omega \cdot \kappa\left(\omega\right) = \frac{2}{\pi} \cdot \mathbb{K}\left(\frac{1}{\omega^2}\right) \quad , \quad \omega^2 > 1. \nonumber
\end{IEEEeqnarray} \\[6pt]
The fluctuation equations for the transverse directions $Y^\mu = X^i = 0$ (\ref{FluctuationEquations5}--\ref{FluctuationEquations6}) can be transformed to the Jacobi form of Lam\'{e}'s equation \cite{Ince26, WhittakerWatson27, Erdelyietal55, MagnusWinkler66, NISTHandbook10}:\\
\begin{IEEEeqnarray}{l}
\frac{d^2z}{du^2} + \left[h - \nu\left(\nu + 1\right) k^2 sn^2\left(u | k^2\right)\right] z = 0, \label{LameEquationI}
\end{IEEEeqnarray}\\
so long as we set,\\
\begin{IEEEeqnarray}{l}
z = \widetilde{y}^\mu(\sigma) \; , \; u = \kappa \omega \left(\sigma + \frac{\pi}{2}\right) \; , \; h = \left(\frac{r}{\kappa \omega}\right)^2 \; , \; \nu\left(\nu + 1\right) = 2 \left(2 \, m^2 + 1\right) \; , \; k = \frac{1}{\omega} \nonumber \\[6pt]
z = \widetilde{x}^i(\sigma) \; , \; u = \kappa \omega \left(\sigma + \frac{\pi}{2}\right) \; , \; h = \left(\frac{r}{\kappa \omega}\right)^2 \; , \; \nu\left(\nu + 1\right) = 4 \left(m^2 - 1\right) \; , \; k = \frac{1}{\omega}. \nonumber
\end{IEEEeqnarray}
\newpage
\subsection[Pulsating Stringy Membranes.]{Pulsating Stringy Membranes. \\}
In order to study the fluctuations of pulsating stringy membranes we set in \ref{FluctuationEquationsY}--\ref{FluctuationEquationsX}: \\
\begin{IEEEeqnarray}{l}
\delta Y^\mu = \sum_{m,n} e^{i n \sigma + i m \delta} \; \widetilde{y}_{m,n}^{\,\mu}\left(\tau\right)\,, \qquad \delta X^i = \sum_{m,n} e^{i n \sigma + i m \delta} \; \widetilde{x}_{m,n}^{\,i}\left(\tau\right)\,, \qquad m \in \mathbb{Z}. \label{FluctuationAnsatzII}
\end{IEEEeqnarray} \\[6pt]
The transverse fluctuation equations \ref{FluctuationEquationsY}--\ref{FluctuationEquationsX} then take the following form (again we omit, for simplicity, the dependencies of $\widetilde{y}_{n,m}^{\,\mu}\left(\tau\right)$ and $\widetilde{x}_{n,m}^{\,i}\left(\tau\right)$ on n, m and $\tau$): \\
\begin{IEEEeqnarray}{c}
\ddot{\widetilde{y}}^{\,\mu} + \left(n^2 + m^2 \widetilde{\Lambda}_0 - \Lambda_0\right) \widetilde{y}^{\,\mu} = 0 \label{FluctuationEquations8}\\[12pt]
\ddot{\widetilde{x}}^{\,i} + \left(n^2 + m^2 \widetilde{\Lambda}_0 - \widetilde{\Lambda}_0\right) \widetilde{x}^{\,i} = 0. \label{FluctuationEquations9}
\end{IEEEeqnarray} \\[6pt]
\begin{figure}
\centering
\includegraphics[scale=0.4]{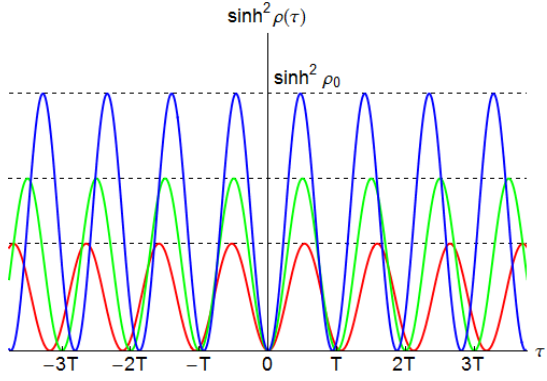}
\caption{Lam\'{e} potential \ref{LamePotentialIII} of stringy membrane \ref{III}--\ref{IIIa}.} \label{Graph:LamePotentialIII}
\end{figure}
For the AdS$_7 \times S^4$ pulsating configuration \ref{III} ($\ell = 2$),\\
\small{\begin{IEEEeqnarray}{c}
Y_0^\mu = 2 \left(\cosh\rho(\tau) \cos t\left(\tau\right) \,,\, 0 \,,\, 0 \,,\, \sinh\rho(\tau) \cos \sigma \,,\, 0 \,,\, \sinh\rho(\tau) \sin \sigma \,,\, 0 \,,\, \cosh\rho(\tau) \sin t\left(\tau\right) \right), \qquad \label{IIIa}
\end{IEEEeqnarray}} \\
\normalsize
we obtain the following Lam\'{e} potential, by solving the equations of motion \ref{ZerothOrderEquation1}--\ref{ZerothOrderEquation2}: \\
\begin{IEEEeqnarray}{c}
\sinh^2\rho\left(\tau\right) = \sinh^2\rho_0 \cdot sn^2\left[\tau \cdot \cosh\rho_0 \, \Big| \, - \tanh^2\rho_0\right], \label{LamePotentialIII}
\end{IEEEeqnarray} \\[6pt]
where $\rho_0$ is given by $4e^2 = \sinh^2 2\rho_0$ and e is a constant of motion (see equation \ref{GKPIIIConstant}). In figure \ref{Graph:LamePotentialIII} we have plotted the potential \ref{LamePotentialIII} for various values of $\rho_0$. The corresponding Lagrange multipliers are given by,
\begin{IEEEeqnarray}{c}
\Lambda_0 = - 2 \sinh^2\rho \quad \& \quad \widetilde{\Lambda}_0 = 4 \sinh^2\rho.
\end{IEEEeqnarray}
The fluctuations along the transverse directions $Y^\mu = X^i = 0$, \ref{FluctuationEquations8}--\ref{FluctuationEquations9}, can again be seen to obey Lam\'{e}'s equation \ref{LameEquationI}. In order to obtain the Jacobi form of the latter we write the potential \ref{LamePotentialIII} as, \\
\begin{IEEEeqnarray}{ll}
\sinh^2\rho\left(\tau\right) = \sinh^2\rho_0 \cdot \left(1 - sn^2\left[\tau \cdot \sqrt{\cosh2\rho_0} + \mathbb{K}\left(\frac{\sinh^2\rho_0}{\cosh2\rho_0}\right) \, \Big| \, \frac{\sinh^2\rho_0}{\cosh2\rho_0}\right]\right) \quad
\end{IEEEeqnarray} \\[6pt]
and make the following substitutions in \ref{LameEquationI}: $u = \tau \cdot \sqrt{\cosh2\rho_0} + \mathbb{K}\left(k^2\right)$ and \\[6pt]
\begin{IEEEeqnarray}{l}
z = \widetilde{y}^\mu(\tau) \; , \; h = \frac{n^2}{\cosh2\rho_0} + 2 k^2 \left(2 \, m^2 + 1\right) \; , \; \nu\left(\nu + 1\right) = 4 m^2 + 2 \; , \; k = \frac{\sinh\rho_0}{\sqrt{\cosh2\rho_0}} \nonumber \\[6pt]
z = \widetilde{x}^i(\tau) \; , \; h = \frac{n^2}{\cosh2\rho_0} + 4 k^2 \left(m^2 - 1\right) \; , \; \nu\left(\nu + 1\right) = 4 m^2 - 4 \; , \; k = \frac{\sinh\rho_0}{\sqrt{\cosh2\rho_0}}. \nonumber
\end{IEEEeqnarray} \\[6pt]
\begin{figure}
\centering
\includegraphics[scale=0.4]{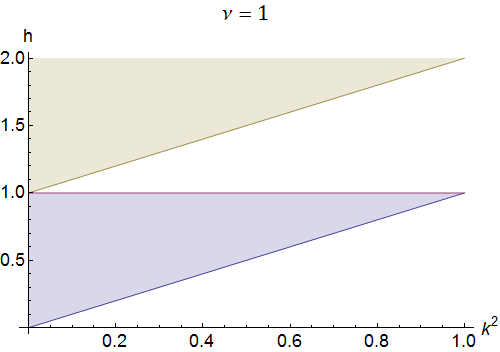} \quad \includegraphics[scale=0.4]{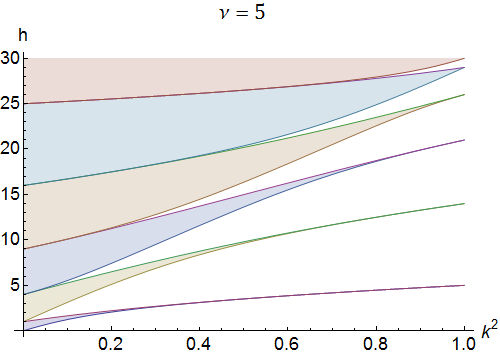}
\caption{Stability bands (colored) of Lam\'{e} equation \ref{LameEquationI} for $\nu = 1$ (left) and $\nu = 5$ (right).} \label{Graph:LameBands}
\end{figure}
Let us now summarize and discuss our results: we have analyzed the fluctuations of the AdS$_7 \times S^4$ stringy membranes \ref{I}, \ref{III} along their transverse directions $Y_0^\mu = X_0^i = 0$ and have found that they fall under Lam\'{e}'s equation \ref{LameEquationI}. For $\nu\left(\nu + 1\right) \in \mathbb{R}$ and $0 < k < 1$, equation \ref{LameEquationI} always has an infinite set of real eigenvalues $a_\nu^s\left(k^2\right)$ and $b_\nu^s\left(k^2\right)$ that correspond to periodic eigenfunctions.\footnote{Note also that Lam\'{e}'s equation \ref{LameEquationI} is symmetric under the exchange $\nu \leftrightarrow - \nu - 1$, so that we only need to consider $\nu > -1/2$ and $\nu(\nu + 1) > -1/4$.} These eigenvalues can be classified into four groups, according to the parity (even or odd) and period (equal to $2\mathbb{K}$ or $4\mathbb{K}$) of their corresponding eigenfunctions (see appendix \ref{LameAppendix}). For a generic eigenvalue h (not necessarily of a periodic eigenfunction), Lam\'{e}'s equation \ref{LameEquationI} is stable iff all corresponding eigenfunctions $z\left(u,h\right)$ are bounded, otherwise it is unstable. It turns out that the intervals of stability are determined by the eigenvalues of periodic solutions:
\begin{equation}
\bcontraction{(a_\nu^0 \,,\, }{a}{_\nu^1) \; \cup \; (}{b} \bcontraction{(a_\nu^0 \,,\, a_\nu^1) \; \cup \; (b_\nu^1 \,,\, }{b}{_\nu^2) \; \cup \; (}{a} \bcontraction{(a_\nu^0 \,,\, a_\nu^1) \; \cup \; (b_\nu^1 \,,\, b_\nu^2) \; \cup \; (a_\nu^2 \,,\, }{a}{_\nu^3) \; \cup \; (}{b} \bcontraction{(a_\nu^0 \,,\, a_\nu^1) \; \cup \; (b_\nu^1 \,,\, b_\nu^2) \; \cup \; (a_\nu^2 \,,\, a_\nu^3) \; \cup \; (b_\nu^3 \,,\, }{b}{_\nu^4) \; \ldots}{}
(a_\nu^0 \,,\, a_\nu^1) \; \cup \; (b_\nu^1 \,,\, b_\nu^2) \; \cup \; (a_\nu^2 \,,\, a_\nu^3) \; \cup \; (b_\nu^3 \,,\, b_\nu^4) \; \cup \; \ldots
\end{equation}
\newpage
Solutions of Lam\'{e}'s equation are stable within the above intervals and unstable outside them. The contractions imply that the relative order of the corresponding endpoints is not a priori known and may thus be reversed, for different values of $\nu \in \mathbb{R}$, $s = 0 \,, 1 \,, 2 \,, \ldots$ and $k \in \left(0\,,1\right)$. Another interesting property of Lam\'{e} eigenvalues is known as "coexistence". In short, coexistence implies that $\nu \in \mathbb{N}$ iff Lam\'{e}'s equation has exactly $\nu + 1$ intervals of stability (bands), following  exactly $\nu + 1$ intervals of instability (gaps). See figure \ref{Graph:LameBands} for plots of the Lam\'{e} bands (colored) and gaps (white) for $\nu = 1$ and $\nu = 5$. \\[6pt]
\begin{table}
\small\begin{IEEEeqnarray}{cccccc}
\text{Ansatz} & u & k & h & z & \nu\left(\nu + 1\right) \nonumber \\[6pt]
\hline \nonumber \\[6pt]
\begin{array}{c} \text{\ref{I}} \\ \text{AdS}_7 \times S^4 \end{array} & \kappa \omega \left(\sigma + \frac{\pi}{2}\right) & \frac{1}{\omega} & \left(\frac{r}{\kappa \omega}\right)^2 & \begin{array}{c} \widetilde{y} \\[12pt] \widetilde{x} \end{array} & \begin{array}{c}  4 \, m^2 + 2 \\[12pt] 4 \left(m^2 - 1\right) \end{array} \nonumber \\[6pt]
\hline \nonumber \\[6pt]
\begin{array}{c} \text{\ref{I}} \\ \text{AdS}_4 \times S^7 \end{array} & \kappa \omega \left(\sigma + \frac{\pi}{2}\right) & \frac{1}{\omega} & \left(\frac{r}{\kappa \omega}\right)^2 & \begin{array}{c} \widetilde{y} \\[12pt] \widetilde{x} \end{array} & \begin{array}{c}  m^2 / 4 + 2 \\[12pt] \frac{1}{4} \left(m^2 - 1\right) \end{array} \nonumber \\[6pt]
\hline \nonumber \\[6pt]
\begin{array}{c} \text{\ref{III}} \\ \text{AdS}_7 \times S^4 \end{array} \; & \; \tau \cdot \sqrt{\cosh2\rho_0} + \mathbb{K}\left(k^2\right) \; & \; \frac{\sinh\rho_0}{\sqrt{\cosh2\rho_0}} \;\; & \;\; \begin{array}{c} \frac{n^2}{\cosh2\rho_0} + k^2 \left(4 \, m^2 + 2\right) \\[12pt] \frac{n^2}{\cosh2\rho_0} + 4 k^2 \left(m^2 - 1\right) \end{array} \;\; & \;\; \begin{array}{c} \widetilde{y} \\[12pt] \widetilde{x} \end{array} \; & \; \begin{array}{c}  4 \, m^2 + 2 \\[12pt] 4 \left(m^2 - 1\right) \end{array} \nonumber \\[6pt]
\hline \nonumber \\[6pt]
\begin{array}{c} \text{\ref{III}} \\ \text{AdS}_4 \times S^7 \end{array} \; & \; \tau \cdot \sqrt{\cosh2\rho_0} + \mathbb{K}\left(k^2\right) \; & \; \frac{\sinh\rho_0}{\sqrt{\cosh2\rho_0}} \;\; & \;\; \begin{array}{c} \frac{n^2}{\cosh2\rho_0} + k^2 \left(m^2 / 4 + 2\right) \\[12pt] \frac{n^2}{\cosh2\rho_0} + \frac{k^2}{4} \left(m^2 - 1\right) \end{array} \;\; & \;\; \begin{array}{c} \widetilde{y} \\[12pt] \widetilde{x} \end{array} \; & \; \begin{array}{c}  m^2 / 4 + 2 \\[12pt] \frac{1}{4} \left(m^2 - 1\right) \end{array} \nonumber \\[6pt]
\hline \nonumber
\end{IEEEeqnarray} \normalsize
\caption{Lam\'{e} fluctuation parameters \ref{LameEquationI} for stringy membranes \ref{I} and \ref{III} in AdS$_{7/4} \times S^{4/7}$.} \label{LameParameters}
\end{table}
We thus see that the stability of Lam\'{e} solutions is organized in (stable) bands and (unstable) gaps. The parameters of Lam\'{e}'s equation \ref{LameEquationI} for each of the examined ans\"{a}tze, are given in table \ref{LameParameters} (the definitions of m, r and n can be found in \ref{FluctuationAnsatzI} and \ref{FluctuationAnsatzII}). The AdS$_7 \times S^4$ results may be easily extended to the AdS$_4 \times S^7$ case (where $R = 2 \ell = 1$ and $\Lambda_0 = -8 \widetilde{\Lambda}_0$) and we have included these as well. The first row of each entry in table \ref{LameParameters} corresponds to the AdS$_{7/4}$ fluctuations $\widetilde{y} \equiv \left\{\widetilde{y}_{r,m}^{\,\mu}\left(\sigma\right) \, , \, \widetilde{y}_{m,n}^{\,\mu}\left(\tau\right)\right\}$ and the second row to the fluctuations on the $S^{4/7}$, $\widetilde{x} \equiv \left\{ \widetilde{x}_{r,m}^{\,i}\left(\sigma\right) \, , \, \widetilde{x}_{m,n}^{\,i}\left(\tau\right)\right\}$. Given $\omega$, $\rho_0$, and $m \in \mathbb{Z}$ ($\kappa = \kappa\left(\omega\right) = 2 / \pi \omega \cdot \mathbb{K}\left(1 / \omega^2\right)$), the allowed values of $r,\, n \in \mathbb{R}$ can be determined in each case by the overlap of the $\widetilde{y}$- and $\widetilde{x}$-bands, the lowest endpoint of which satisfies:
\begin{IEEEeqnarray}{ll}
h_{\text{min}} \geq 0 \text{, in ansatz \ref{I}} \ \& \ & h_{\text{min}} \geq \left(4 \, m^2 + 2\right) \, \frac{\sinh^2\rho_0}{\cosh2\rho_0} \text{, in ansatz \ref{III} (AdS}_7 \times S^4 \text{)} \nonumber \\[6pt]
& h_{\text{min}} \geq \left(m^2 / 4 + 2\right) \, \frac{\sinh^2\rho_0}{\cosh2\rho_0} \text{, in ansatz \ref{III} (AdS}_4 \times S^7 \text{)}. \qquad \quad
\end{IEEEeqnarray}
\section[Discussion.]{Discussion.\label{Discussion}}

In this paper, we have studied the stringy properties of uncharged bosonic membranes in AdS$_7 \times S^4$ and AdS$_4 \times S^7 / \mathbb{Z}_k$. We have examined the conditions under which the string sigma model in AdS$_5 \times S^5$ may be embedded in the membrane sigma model in AdS$_{4,7} \times S^{7,4}$. Specifically, we have found that all string configurations of AdS$_5$ may be reproduced by membranes living in AdS$_7 \times S^4$. Moreover, all string solitons that live in $AdS_4 \subset AdS_5$ may be reproduced by membranes of AdS$_4 \times S^7$. We have also shown how logarithmic scaling violations (i.e. $E - S \sim \ln S$) for membranes living in AdS$_4 \times S^7 / \mathbb{Z}_k$ may be obtained, generalizing the work of Hartnoll and Nu\~{n}ez \cite{HartnollNunez02}.  \\[6pt]
There's absolutely no magic in obtaining stringy behavior from membranes on AdS$_m \times S^n$. The corresponding setups are essentially one-dimensional in each of the two product spaces, having no dynamics in one of them (the n-sphere). Viewed together as an ensemble, they have two independent dimensions. Hence their membrane nature. Our treatment is very similar to that of Duff-Howe-Inami-Stelle \cite{DuffHoweInamiStelle87}, albeit with a different motivation \cite{HartnollNunez02, AxenidesFloratos07, BruguesRojoRusso04}. Compared to \cite{DuffHoweInamiStelle87}, and apart from considering only bosonic membranes in an $AdS_m \times S^n$ background (i.e. a product of two manifolds), we haven't actually performed a double dimensional reduction (as e.g. in \cite{BruguesRojoRusso04}), although it may have seemed so. In this work, we have been primarily interested in the applications of the GKP method. A posteriori, analogous string-membrane reductions could be found in \cite{Kamani03a, Kamani03b, Gangopadhyay07}. \\[6pt]
Secondly, we have analyzed the stability of stringy membranes in the linearized approximation. We have demonstrated that the similarities between stringy membranes and strings cannot be extended beyond leading order, since the perturbation equations depend on the second world-volume coordinate $\delta$, which cannot be eliminated from the equations themselves. By studying the stability of stringy membranes along their transverse directions we have found that they are governed by Lam\'{e}'s equation. Therefore, they typically exhibit the standard stability/instability pattern of bands and gaps. Interestingly (for $m = 0$ in table \ref{LameParameters}), our analysis recovers the single-band/single-gap structure of the AdS$_3$ string case \cite{Beccariaetal10a, Beccariaetal10b}. At this point, important issues of interpretation arise for both strings and membranes. Firstly, does the Lam\'{e} band/gap structure that anti-de Sitter strings and membranes possess, admit a particle interpretation? Moreover, what is the holographic dual of the Lam\'{e} instability phenomenon in question? In what follows, we conclude our work with a detailed exposition of our results as well as some prospects for further work on open issues that emerge from them. \\[12pt]
$\bullet$ \textit{Scaling dimensions and stringy membranes}. \\[6pt]
Solution \ref{I} essentially coincides with the $AdS_4 \times S^7$, "type-I" solution of Hartnoll and Nu\~{n}ez \cite{HartnollNunez02}, although it is written in terms of the Polyakov action on $AdS_7 \times S^4$ (see section \ref{AdS4xS7}, for $AdS_4 \times S^7$).  As it is well-known, the folded closed string of AdS$_3$ is dual to the operator Tr$\left[\mathcal{Z} \, \mathcal{D}_+^S \, \mathcal{Z}\right] + \ldots$ of the SL(2) sector of $\mathcal{N} =4$, SYM.\footnote{$\mathcal{Z}$, $\mathcal{W}$, $\mathcal{Y}$ are the three complex scalars of $\mathcal{N} =4$, SYM, composed out of its six real scalars $\Phi$. Also $\mathcal{D}_+ = \mathcal{D}_0 + \mathcal{D}_3$, $\mathcal{D}_- = \mathcal{D}_1 + \mathcal{D}_2$, denote the light-cone derivatives. The dots in a trace operator generally stand for terms that are built by permuting trace fields $\mathcal{Z}$ and impurities, $\mathcal{W}$, $\mathcal{Y}$, $\mathcal{D}_\pm$.} Therefore, as postulated in \cite{HartnollNunez02} and in complete analogy with the stringy case \cite{GubserKlebanovPolyakov02}, this membrane configuration is expected to correspond to twist-2 gauge theory operators, with anomalous scaling dimensions given, at strong coupling, by the equations of the corresponding leading Regge trajectories (writing S for the charge $S_1 = S^{12}$ in \ref{MembraneCyclicCharge2} and defining $\sqrt{\lambda'} \equiv R \, \ell^2 / g_s \ell_s^3$): \\
\begin{IEEEeqnarray}{l}
E^2 = 2 \, \sqrt{\lambda'} \, S + \ldots \quad \left(\text{Short Stringy Membranes, } S \ll \sqrt{\lambda'}\right) \\[6pt]
E - S = f\left(\lambda'\right) \ln\frac{S}{\sqrt{\lambda'}} + \ldots \quad \left(\text{Long Stringy Membranes, } S \gg \sqrt{\lambda'}\right).
\end{IEEEeqnarray} \\
At the classical level, it is rather easy to obtain the full "short" series whereas finding the "long" series presents more challenges. A method that potentially generates all of the subleading "long" terms was presented in \cite{GeorgiouSavvidy11}, along with the proof of a formula that links the expressions for the anomalous dimensions, in the "short" and the "long" regimes (see also appendix \ref{Appendix:StringActions}). What is more, the long series was found to satisfy the Moch-Vermaseren-Vogt (MVV) constraints that follow from a "reciprocity", aka "parity-preserving" relation. Originally proposed by Gribov and Lipatov \cite{GribovLipatov72} in the context of deep inelastic scattering (DIS), the so-called "strong" reciprocity relation has been verified for twist-two operators, up to three loops in perturbative QCD \cite{BassoKorchemsky06} and up to four loops in weakly coupled, $\mathcal{N} = 4$, SYM \cite{BeccariaForini09, BeccariaForiniMacorini10}. It was claimed in \cite{GeorgiouSavvidy11} that reciprocity is very likely satisfied by twist-two operators in string perturbation theory as well. As we have just seen, all of these statements naturally carry over to stringy membranes. \\[6pt]
On the other hand it is known that the "cusp anomalous dimension" $f(\lambda)$ receives quantum corrections that are calculated in superstring theory by evaluating the Lam\'{e} fluctuation determinants \cite{Beccariaetal10a}. Since the quadratic supermembrane sigma model on AdS$_{7/4} \times S^{4/7}$ is completely different from the corresponding model of superstrings (we have seen an instance of this in the fluctuations of stringy membranes), we expect that the quantum corrections to the anomalous dimensions of twist-2 operators, as calculated from AdS$_{7/4} \times S^{4/7}$ supermembranes, will differ from the purely stringy ones. \\[12pt]
$\bullet$ \textit{Integrability}. \\[6pt]
The equations of motion and all constraint equations of stringy membranes \ref{ZerothOrderEquation1}--\ref{ZerothOrderEquation2} are identical to the corresponding equations of strings that rotate in anti-de Sitter space. As such, they may be shown to be equivalent to the generalized sinh-Gordon equation through a reduction of the Pohlmeyer type \cite{Pohlmeyer75}. In complete analogy with AdS strings, stringy membranes in AdS$_{(2,3,4)}$ thus turn out to be equivalent to the Liouville, sinh-Gordon and B$_2$-Toda model respectively (cf. \cite{BarbashovNesterenko81, deVegaSanchez92, LarsenSanchez96}). \\[6pt]
Another outcome of our analysis concerns the dual gauge theories. The generalization of the GKP method to theories which contain extended objects other than strings, offers a method to compare their dual CFTs by means of studying the integrable sectors that they probably share in the bulk. That is, useful insights about the behavior of one theory can be extracted by studying a similar sector of the other, despite the fact that the theories might significantly differ or even have different dimensionalities. In the present work, the following dualities that contain states/operators for which $\Delta - S \sim \ln S$ were examined: \\
{\renewcommand{\arraystretch}{1.5} \renewcommand{\tabcolsep}{0.2cm} \begin{center} \begin{tabular}{|c|c|}
\cline{1-2}
{\color{blue}{Gauge Theory}} & {\color{blue}{dual Gravity Theory}} \\
\cline{1-2}
$\mathcal{N} = 4$,  $SU\left(N\right)$, Super Y-M Theory & IIB String Theory on $AdS_5 \times S^5$ \\
$\mathcal{N} = 8$ SCFT \Big/ $A_{N - 1} \left(2,0\right)$ SCFT & M-Theory on $AdS_{4/7} \times S^{7/4}$   \\
\cline{1-2}
$\mathcal{N} = 6$,  $U\left(N\right)_k \times U\left(N\right)_{-k}$, Super C-S Theory & \\
$N\rightarrow\infty$ & M-Theory on $AdS_4 \times S^7/\mathbb{Z}_k$ \\
$k^5 \gg N \rightarrow \infty$, $\lambda \equiv 2\pi^2N/k =$ const. & IIA String Theory on $AdS_4 \times \text{CP}^3$ \\
\cline{1-2}
\end{tabular} \\[24pt] \end{center}}
We believe that our study of stringy membranes strengthens the conjecture put forward by Bozhilov in \cite{Bozhilov07a} that (a) $\mathcal{N} = 4$, $SU(N)$ SYM theory (dual to IIB String Theory on $AdS_5 \times S^5$), (b) $A_{N - 1}(2,0)$, SCFT (dual to M-theory on $AdS_7 \times S^4$) and (c) $\mathcal{N} = 8$, SCFT (dual to M-theory on $AdS_4 \times S^7$) all possess common integrable sectors, and we would also like to guess that this "AdS family" could contain more members (e.g. QCD, $\mathcal{N} = 6$, quiver Super Chern-Simons \cite{ABJM08}, $\mathcal{N} = 1$, SYM \cite{HartnollNunez02, Acharya00, AtiyahMaldacenaVafa01, AtiyahWitten01, CveticGibbonsLuPope01, Gukov03}, etc.). In another---yet similar---direction, it has been shown in \cite{BelitskyDerkachovKorchemskyManashov04} that $\mathcal{N} = 0 \,, 1 \,, 2 \,, 4$, SYM theories all possess a common universal one-loop dilatation operator. \\[12pt]
$\bullet$ \textit{Possible generalizations}. \\[6pt]
We couldn't think of a more general argument showing that all (super-) string theories that can be formulated on $\text{AdS}_5$ and their dual gauge sectors, are included in an $AdS_7 \times S^4$ (super-) membrane theory and its dual SCFT respectively. Moreover, double dimensional reduction \cite{DuffHoweInamiStelle87}, doesn't generally work in the case,

\begin{center} $\left\{\text{membranes/}AdS_{4,7} \times S^{7,4}\right\} \longrightarrow \left\{\text{strings/}AdS_5 \times S^5\right\}$, \end{center}

thus we have no a priori reason to expect that string theory on $AdS_5 \times S^5$ is contained in M-theory on $AdS_{4,7} \times S^{7,4}$. It would, nevertheless, be extremely interesting to investigate the extent up to which the results of Duff-Howe-Inami-Stelle \cite{DuffHoweInamiStelle87} can be applied to the $AdS_{4,7} \times S^{7,4}$ case as well. That is find out which embeddings of the full Green-Schwarz action on $AdS_5 \times S^5$ \cite{MetsaevTseytlin98, KalloshRahmfeld98, KalloshRahmfeldRajaraman98, KalloshTseytlin98, DrukkerGrossTseytlin00, Tseytlin00}, into the full supermembrane action on $AdS_{4,7} \times S^{7,4}$ \cite{DallAgataFabbriFraserFreTermoniaTrigiante98, deWitPeetersPlefkaSevrin98, Claus98, PastiSorokinTonin98} are allowed, much along the (bosonic) lines of the present paper. \\[6pt]
Going further, one could attempt to study the difference of the membrane and string Polyakov actions, $\mathcal{S}_2 - \mathcal{S}_1$, in more complex setups. Similarly, one could prove that any membrane soliton may be obtained by going to higher-dimensional extended objects (e.g. a 3- or a 5-brane) living in more spacetime dimensions. In general one could claim that any p-brane soliton, living in pure $\text{AdS}_m$, may be obtained from a $(p + 1)$-brane living in $AdS_{m'} \times S^{m + n + 1 - m'}$ or a $(p + q)$-brane living in an adequately generalized spacetime. \\[6pt]
\section{Acknowledgements.}
We thank G. Georgiou and S. Nicolis for illuminating discussions. We are also grateful to M. Alishahiha, I. Bakas and D.Sorokin for patiently explaining their work to us. \\[6pt]
The research of E.F. is implemented under the "ARISTEIA" action (Code no.1612, D.654) and title "Hologaphic Hydrodynamics" of the "operational programme education and lifelong learning" and is co-funded by the European Social Fund (ESF) and National Resources.  The research of M.A. and G.L. is supported in part by the Operational Programme "Competitiveness and Entrepreneurship" (OPCE II), Action "Development Proposals of Research Organizations - KRIPIS", NSRF 2007-2013. \\[6pt]
G.L. kindly acknowledges hospitality at APCTP during the program "Holography 2013: Gauge/gravity duality and strongly correlated systems". \\[6pt]
\appendix
\section[Spinning Strings in $AdS_5 \times S^5$.]{Spinning Strings in $AdS_5 \times S^5$.\label{Appendix:StringActions}}
In this appendix we will sketch the rudiments of the Gubser-Klebanov-Polyakov (GKP) string configurations \cite{GubserKlebanovPolyakov02} that we discussed in our paper. First consider a closed and uncharged bosonic string in $AdS_5\times S^5$, in the global coordinate system: \\
\begin{IEEEeqnarray}{lcl}
Y_0 + i Y_5 = \ell \cosh\rho \, e^{it} && X_1 + i X_2 = R \cos\overline{\theta}_1 \, e^{i \overline{\phi}_1} \nonumber \\[6pt]
Y_1 + i Y_2 = \ell \sinh\rho \cos\theta \, e^{i\phi_1} & \quad \& \quad & X_3 + i X_4 = R \sin\overline{\theta}_1 \cos\overline{\theta}_2 \, e^{i \overline{\phi}_2} \\[6pt]
Y_3 + i Y_4 = \ell \sinh\rho \sin\theta \, e^{i\phi_2} && X_5 + i X_6 = R \sin\overline{\theta}_1 \sin\overline{\theta}_2 \, e^{i \overline{\phi}_3}, \nonumber
\end{IEEEeqnarray} \\
where $Y^\mu$ and $X^i$ are the embedding coordinates of AdS$_5 \times S^5$ (see \ref{EmbeddingCoordinates1}--\ref{EmbeddingCoordinates2}) and $\rho \geq 0, \, t \in \left[0, \, 2\pi\right), \, \overline{\theta}_1 \in \left[0, \, \pi\right], \, \theta, \, \phi_1, \, \phi_2, \, \overline{\theta}_2, \, \overline{\phi}_1, \, \overline{\phi}_2, \, \overline{\phi}_3 \in \left[0, \, 2\pi\right)$. The string Polyakov action in the conformal gauge ($\gamma_{ab} = \eta_{ab}$) is given by:\footnote{$T_1$ is the string tension, $T_1 \equiv 1/2\pi\alpha'$.} \\
\begin{IEEEeqnarray}{ll}
\mathcal{S} & = - \frac{T_1}{2} \int \sqrt{-\gamma} \, \gamma_{ab} \Big[G_{mn}^{AdS}(y) \partial^a y^m \partial^b y^n + \, G_{mn}^S(x) \partial^a x^m \partial^b x^n\Big] d\tau \, d\sigma = \nonumber \\[6pt]
& = \frac{T_1}{2} \int \Big[G_{mn}^{AdS}(y) \left(\dot{y}^m \dot{y}^n - y'\,^m y'\,^n\right) + G_{mn}^S(x)\left(\dot{x}^m \dot{x}^n - x'\,^m x'\,^n\right)\Big] d\tau \, d\sigma, \label{StringPolyakov}
\end{IEEEeqnarray}
\newpage
where $y^m \equiv \left(t, \, \rho, \, \theta, \, \phi_1, \, \phi_2\right)$ and $x^m \equiv \left(\overline{\theta}_1, \, \overline{\theta}_2, \, \overline{\phi}_1, \, \overline{\phi}_2, \, \overline{\phi}_3\right)$. Our first configuration consists of a folded closed string that rotates at the equator of $S^3$ of $AdS_5$: \\
\begin{equation}
\Big\{t = \kappa \tau, \, \rho = \rho(\sigma), \, \theta = \kappa \omega \tau, \, \phi_1 = \phi_2 = 0\Big\} \times \Big\{\overline{\theta}_1 = \overline{\theta}_2 = \overline{\phi}_1 = \overline{\phi}_2 = \overline{\phi}_3 = 0\Big\}. \label{GKPAnsatzI} \\[12pt]
\end{equation}
In embedding coordinates this ansatz reads: \\
\begin{IEEEeqnarray}{lcl}
Y_0 = \ell \cosh\rho(\sigma) \cos\kappa\tau & \qquad , \qquad & X_1 = R = \ell \nonumber \\
Y_1 = \ell \sinh\rho(\sigma) \cos\kappa\omega\tau && X_2 = X_3 = X_4 = X_5 = X_6 = 0 \nonumber \\
Y_2 = \ell \sinh\rho(\sigma) \sin\kappa\omega\tau && \\
Y_3 = Y_4 = 0 \nonumber \\
Y_5 = \ell \cosh\rho(\sigma) \sin\kappa\tau. \nonumber
\end{IEEEeqnarray} \\
Its equations of motion and Virasoro constraints become: \\
\begin{IEEEeqnarray}{l}
\rho'' + \kappa^2 \left(\omega^2 - 1\right) \sinh\rho \cosh\rho = 0  \label{GKPEquationI}\\[6pt]
\rho'\,^2 - \kappa^2\left(\cosh^2\rho - \omega^2 \sinh^2\rho\right) = 0. \label{GKPConstraintI}
\end{IEEEeqnarray} \\
Depending on the value of the angular velocity $\omega$, two basic cases are obtained: \\[12pt]
(i). $\omega^2 > 1$ : A folded closed rigidly rotating string, with cusps at $d\sigma / d\rho \Big|_{\rho_0} = \infty$,
\begin{equation}
0 \leq \sinh^2\rho \leq \sinh^2\rho_0 = \frac{1}{\omega^2 - 1} = q < \infty. \nonumber \\[6pt]
\end{equation}

\hspace*{0.5cm} a. "Short" Strings: $\omega \rightarrow \infty$ \,, $\rho_0 \sim 1/\omega$. \\[6pt]
\hspace*{0.5cm} b. "Long" Strings: $\omega = 1 + 2\eta \rightarrow 1^+$ \,, $\rho_0 \sim \ln1/n \rightarrow \infty$. \\[6pt]

(ii). $\omega^2 < 1$ : Two oppositely oriented rigidly rotating Wilson loops, with \\
\begin{equation}
0 \leq \sinh^2\rho \leq \sinh^2\rho_0 = \infty. \nonumber \\[6pt]
\end{equation}
This system has two cyclic coordinates---namely t and $\theta$---so that the conservation laws are the following: \\
\begin{IEEEeqnarray}{l}
E = \left|\frac{\partial L}{\partial \dot{t}}\right| = \frac{\ell^2}{2 \pi \alpha'} \int_0^{2\pi} \kappa \,\cosh^2\rho \, d\sigma = 4 \cdot \frac{\ell^2}{2 \pi \alpha'} \int_0^{\rho_0} \frac{\cosh^2\rho \, d\rho}{\sqrt{1 - \left(\omega^2 - 1\right)\sinh^2\rho}} \label{GKPIEnergy1}\\[12pt]
S = \frac{\partial L}{\partial \dot{\theta}} = \frac{\ell^2}{2 \pi \alpha'} \int_0^{2\pi} \kappa \, \omega \, \sinh^2\rho \, d\sigma = 4 \cdot \frac{\ell^2}{2 \pi \alpha'} \int_0^{\rho_0} \frac{\omega \, \sinh^2\rho \, d\rho}{\sqrt{1 - \left(\omega^2 - 1\right)\sinh^2\rho}}. \label{GKPISpin1}
\end{IEEEeqnarray}
\newpage
The string essentially contains four segments extending between $\rho = 0$ and $\rho = \rho_0$ and this accounts for the factor 4 in front of the $\rho$-integrals. One also has to calculate the length of the string,
\begin{IEEEeqnarray}{l}
\sigma \cdot \kappa = \int_0^\rho \frac{d\rho}{\sqrt{1 - \left(\omega^2 - 1\right)\sinh^2\rho}}, \label{GKPILength1}
\end{IEEEeqnarray} \\
where $\kappa$ is a factor needed to fix $\rho(\sigma = \pi/2) = \rho_0$. \\[24pt]
$\blacksquare \quad \underline{\omega^2 > 1}$\,. \\[12pt]
\begin{figure}
\begin{center}
\includegraphics[scale=0.45]{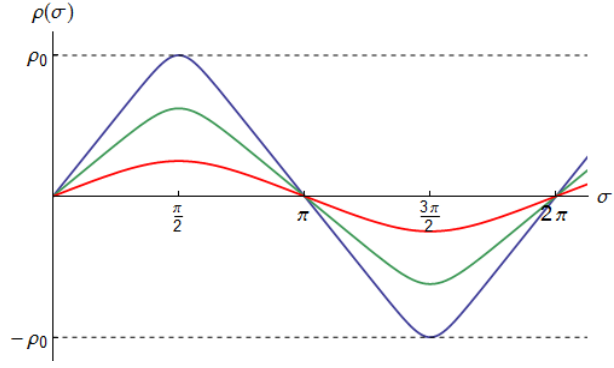}
\caption{$\rho = \rho\left(\sigma\right)$ of the folded closed string \ref{GKPAnsatzI}, when $\omega^2 > 1$.} \label{Graph:StringLengthI}
\end{center}
\end{figure}
For the case (i) of the closed and folded string with $\omega^2 > 1$, it's $\omega \cdot \tanh \rho_0 = 1$ so that the integrals \ref{GKPIEnergy1} - \ref{GKPILength1} take simpler forms and can be expressed in terms of complete elliptic functions:\footnote{Our conventions for the elliptic integrals and elliptic functions follow Abramowitz-Stegun \cite{AbramowitzStegun65}.}

\begin{IEEEeqnarray}{l}
\rho(\sigma) = \text{arctanh} \left[\frac{1}{\omega} sn\left(\kappa \omega \sigma \, \Bigg| \,\frac{1}{\omega^2}\right)\right], \quad \kappa = \frac{2}{\pi\omega} \, \mathbb{K}\left(\frac{1}{\omega^2}\right), \quad \omega = \coth\rho_0 \label{GKPILength2}
\end{IEEEeqnarray}

\begin{IEEEeqnarray}{l}
E(\omega) = \frac{2 \ell^2}{\pi \alpha'} \cdot \frac{\omega}{\omega^2 - 1} \, \mathbb{E} \left(\frac{1}{\omega^2}\right) \label{GKPIEnergy2}
\end{IEEEeqnarray}

\begin{figure}
\begin{center}
\includegraphics[scale=0.45]{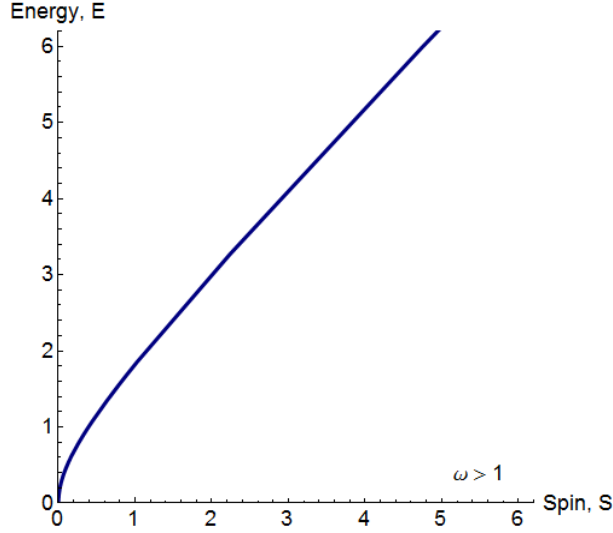}
\caption{Energy and spin of the folded closed string \ref{GKPAnsatzI}, for $\omega^2 > 1$.} \label{Graph:Energy-SpinIb}
\end{center}
\end{figure}

\begin{IEEEeqnarray}{l}
S(\omega) = \frac{2 \ell^2}{\pi \alpha'} \cdot \left(\frac{\omega^2}{\omega^2 - 1} \, \mathbb{E} \left(\frac{1}{\omega^2}\right) - \mathbb{K} \left(\frac{1}{\omega^2}\right)\right). \label{GKPISpin2}
\end{IEEEeqnarray} \\[6pt]
We have plotted $\rho\left(\sigma\right)$ for various values of $\omega$ in figure \ref{Graph:StringLengthI}. In figure \ref{Graph:Energy-SpinIb} we have plotted the energy of the string as a function of its spin, $E = E(S)$.
\newpage
Following \cite{GeorgiouSavvidy11} we may also establish a kind of duality between short and long folded closed strings in AdS$_3$. To begin, there's a known formula between the complete elliptic integrals of the first and second kinds, namely Legendre's relation (see e.g. \cite{AbramowitzStegun65}): \\[6pt]
\begin{equation}
\mathbb{E}(k) \mathbb{K}(k') + \mathbb{K}(k) \mathbb{E}(k') - \mathbb{K}(k) \mathbb{K}(k') = \frac{\pi}{2}, \label{Legendre} \\[12pt]
\end{equation}
where the arguments of the elliptic integrals $k = 1 / \omega^2$ and $k' = 1 / \omega' \, ^2$ satisfy $k + k' = 1$. We thus see that large values of $\omega' \rightarrow \infty$ ("short" strings) correspond to values of $\omega \rightarrow 1^+$ near unity ("long" strings) and \ref{Legendre} then provides a map between the corresponding energies and spins.
Solving \ref{GKPIEnergy2} and \ref{GKPISpin2} for $\mathbb{E}(k)$ and $\mathbb{K}(k)$ and substituting in \ref{Legendre}, we get the following duality relation between classical folded short and long strings: \\[6pt]
\begin{equation}
\frac{1}{\omega} E S' + \frac{1}{\omega'} E' S - S S' = \frac{2 \lambda}{\pi}, \label{short-long1}
\end{equation} \\
where the value of the 't Hooft coupling, $\lambda = \ell^4 / \alpha'\,^2$ has also been used. There's yet another useful expression of \ref{short-long1} in terms of the anomalous dimension $\gamma \equiv E - S$, \\[6pt]
\begin{equation}
\frac{1}{\omega} \gamma S' + \frac{1}{\omega'} \gamma' S + \left(\frac{1}{\omega} + \frac{1}{\omega'} - 1\right) S S' = \frac{2 \lambda}{\pi}. \label{short-long2} \\[12pt]
\end{equation}
The second string configuration that we will examine consists of a closed string that pulsates at the equator of $S^3$ of $AdS_5$: \\
\begin{equation}
\Big\{t = t\left(\tau\right), \, \rho = \rho(\tau), \, \theta = 0, \, \phi_1 = w \sigma, \, \phi_2 = 0\Big\} \times \Big\{\overline{\theta}_1 = \overline{\theta}_2 = \overline{\phi}_1 = \overline{\phi}_2 = \overline{\phi}_3 = 0\Big\}. \label{GKPAnsatzIII}
\end{equation}
\newpage
\begin{figure}
\begin{center}
\includegraphics[scale=0.45]{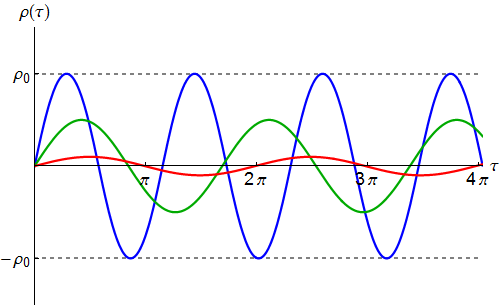}
\caption{$\rho = \rho\left(\tau\right)$ of the pulsating closed string \ref{GKPAnsatzIII}.} \label{Graph:StringLengthIII}
\end{center}
\end{figure}
In AdS$_5 \times S^5$ embedding coordinates the solution reads: \\
\begin{IEEEeqnarray}{lcl}
Y_0 = \ell \cosh\rho(\tau) \cos t(\tau) && X_1 = R = \ell \nonumber \\
Y_1 = \ell \sinh\rho(\tau) \cos w \sigma & \qquad , \qquad & X_2 = X_3 = X_4 = X_5 = X_6 = 0 \nonumber \\
Y_2 = \ell \sinh\rho(\tau) \sin w \sigma && \\
Y_3 = Y_4 = 0 && \nonumber \\
Y_5 = \ell \cosh\rho(\tau) \sin t(\tau). && \nonumber
\end{IEEEeqnarray} \\
The equations of motion and the Virasoro constraints become: \\
\begin{IEEEeqnarray}{l}
\ddot{t} \cosh^2\rho + 2 \, \dot{t} \, \dot{\rho} \, \cosh\rho \sinh\rho = 0 \label{t-equation} \\[6pt]
\ddot{\rho} + \sinh\rho \cosh\rho \left(\dot{t}^2 + w^2\right) = 0 \label{rho-equation} \\[6pt]
\dot{\rho}^2 - \dot{t}^2 \cosh^2\rho + w^2\sinh^2\rho = 0. \label{GKPIII-constraint}
\end{IEEEeqnarray} \\
The conserved energy, as well as the string length are given by: \\
\begin{IEEEeqnarray}{l}
E = \left|\frac{\partial L}{\partial \dot{t}}\right| = \frac{\ell^2}{2 \pi \alpha'} \int_0^{2\pi} \dot{t} \cosh^2\rho \, d\sigma = \frac{\ell^2}{\alpha'} \cdot \dot{t} \, \cosh^2\rho = \frac{w \, e \, \ell^2}{\alpha'} = w \sqrt{\lambda} \, e \label{GKPIIIEnergy} \\[12pt]
\tau \left(\rho\right) = \int_{0}^{\rho} \frac{\cosh\rho \, d\rho}{w \sqrt{e^2 - \cosh^2\rho\sinh^2\rho}} = \int_{0}^{\sinh\rho} \frac{dx}{w \sqrt{e^2 - x^2 - x^4}}. \label{GKPIIILength}
\end{IEEEeqnarray} \\[6pt]
In addition, $\rho < \rho_0$ must hold with $\rho_0$ satisfying \\
\begin{IEEEeqnarray}{c}
e = \frac{\dot{t}}{w} \cdot \cosh^2\rho\left(\tau\right) \equiv \sinh\rho_0 \cosh\rho_0 = \text{const.} \label{GKPIIIConstant}
\end{IEEEeqnarray} \\[6pt]
Performing the integral \ref{GKPIIILength} we obtain $\tau \left(\rho\right)$ and, by inversion, $\rho \left(\tau\right)$: \\
\begin{IEEEeqnarray}{l}
\rho \left(\tau\right) = \text{arcsinh}\left[\sinh\rho_0 \cdot sn \left(w \tau \cosh\rho_0 \Big| -\tanh^2\rho_0\right)\right].
\end{IEEEeqnarray} \\[6pt]
This is an oscillatory time-periodic solution that we have plotted for various $\rho_0$ in figure \ref{Graph:StringLengthIII}.\\[6pt]
\section[Lam\'{e}'s Equation.]{Lam\'{e}'s Equation. \label{LameAppendix}}

\begin{figure}
\centering
\includegraphics[scale=0.3]{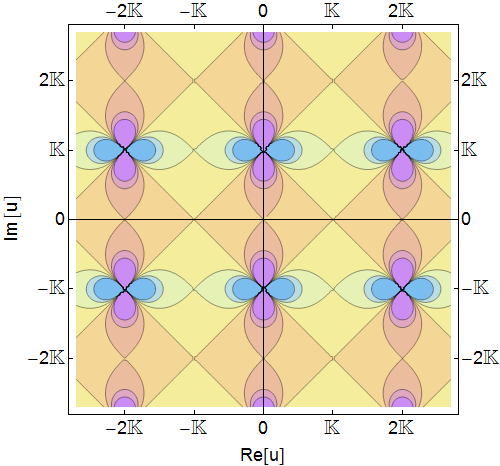} \qquad \includegraphics[scale=0.3]{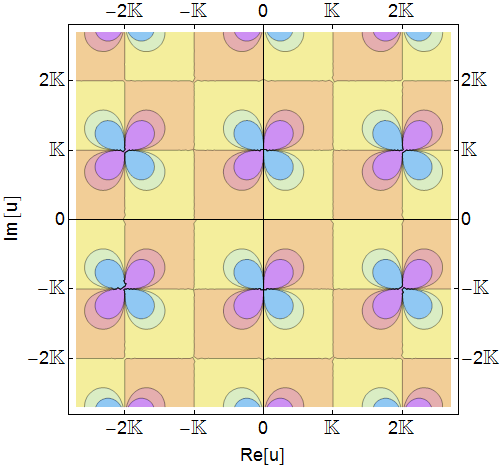}
\caption{Real and imaginary parts of the Lam\'{e} potential, sn$^2\left(u|1/2\right)$.} \label{LamePotentialIV}
\end{figure}

We saw in section \ref{Fluctuations} that the fluctuation equations of stringy membranes \ref{I}--\ref{III}, can be reduced to the Jacobian form of Lam\'{e}'s equation, namely, \\
\begin{IEEEeqnarray}{l}
\frac{d^2z}{du^2} + \left[h - \nu\left(\nu + 1\right) k^2 sn^2\left(u | k^2\right)\right] z = 0, \label{LameEquationIII}
\end{IEEEeqnarray} \\
where $\nu\left(\nu + 1\right) \in \mathbb{R}$ and $0 < k < 1$ \cite{Erdelyietal55, NISTHandbook10}. The potential of Lam\'{e}'s equation, $sn^2\left(u | k^2\right)$ is a doubly periodic function with (primitive) real and imaginary periods equal to $2\mathbb{K}\left(k^2\right)$ and $2 i \mathbb{K}'\left(k^2\right)$ respectively. It is depicted in figure \ref{LamePotentialIV}. The eigenfunctions of Lam\'{e}'s equation (Lam\'{e} functions), that have real periods are: \\
\renewcommand{\arraystretch}{.2}
\renewcommand{\tabcolsep}{0.2cm}
\begin{center}\begin{tabular}{ccccc}
\hline \\
eigenfunction $z\left(u\right)$ & eigenvalue h & parity of $z\left(u\right)$ & parity of $z\left(u - \mathbb{K}\right)$ & period of $z\left(u\right)$ \\[6pt]
\hline \hline \\
${Ec}_\nu^{2n}\left(u,k^2\right)$ & $a_\nu^{2n}\left(k^2\right)$ & even & even & $2\mathbb{K}$ \\[6pt]
\hline \\
${Ec}_\nu^{2n + 1}\left(u,k^2\right)$ & $a_\nu^{2n + 1}\left(k^2\right)$ & odd & even & $4\mathbb{K}$ \\[6pt]
\hline \\
${Es}_\nu^{2n + 1}\left(u,k^2\right)$ & $b_\nu^{2n + 1}\left(k^2\right)$ & even & odd & $4\mathbb{K}$ \\[6pt]
\hline \\
${Es}_\nu^{2n + 2}\left(u,k^2\right)$ & $b_\nu^{2n + 2}\left(k^2\right)$ & odd & odd & $2\mathbb{K}$ \\[6pt]
\hline
\end{tabular} \\[24pt] \end{center}
where $n = 0 \,,\, 1 \,,\, 2 \,,\, \ldots $. The eigenvalues have the following ordering properties \cite{Erdelyietal55, NISTHandbook10}:
\begin{IEEEeqnarray}{ll}
a_\nu^0 < a_\nu^1 < a_\nu^2 < a_\nu^3 \, \ldots \,,\quad & a_\nu^n \rightarrow \infty \text{ as } n \rightarrow \infty \nonumber \\[6pt]
b_\nu^1 < b_\nu^2 < b_\nu^3 < b_\nu^4 \, \ldots \,,\quad & b_\nu^n \rightarrow \infty \text{ as } n \rightarrow \infty  \nonumber \\[6pt]
a_\nu^0 < b_\nu^1 < a_\nu^2 < b_\nu^3 \, \ldots \nonumber \\[6pt]
a_\nu^1 < b_\nu^2 < a_\nu^3 < b_\nu^4 \, \ldots \nonumber
\end{IEEEeqnarray}
The intervals of stability of \ref{LameEquationIII} follow from a theorem known as \textit{oscillation theorem} \cite{MagnusWinkler66}: \\
\begin{equation}
\bcontraction{(a_\nu^0 \,,\, }{a}{_\nu^1) \; \cup \; (}{b} \bcontraction{(a_\nu^0 \,,\, a_\nu^1) \; \cup \; (b_\nu^1 \,,\, }{b}{_\nu^2) \; \cup \; (}{a} \bcontraction{(a_\nu^0 \,,\, a_\nu^1) \; \cup \; (b_\nu^1 \,,\, b_\nu^2) \; \cup \; (a_\nu^2 \,,\, }{a}{_\nu^3) \; \cup \; (}{b} \bcontraction{(a_\nu^0 \,,\, a_\nu^1) \; \cup \; (b_\nu^1 \,,\, b_\nu^2) \; \cup \; (a_\nu^2 \,,\, a_\nu^3) \; \cup \; (b_\nu^3 \,,\, }{b}{_\nu^4) \; \ldots}{}
(a_\nu^0 \,,\, a_\nu^1) \; \cup \; (b_\nu^1 \,,\, b_\nu^2) \; \cup \; (a_\nu^2 \,,\, a_\nu^3) \; \cup \; (b_\nu^3 \,,\, b_\nu^4) \; \cup \; \ldots \label{LameBands} \\[12pt]
\end{equation}
where the contractions have been included to signify that the relative order of the contracted terms is not generally known and can therefore be reversed, for given values of $\nu$ and $k^2$.
 \\
For $\nu \in \mathbb{R}$, the expression $\nu\left(\nu + 1\right) \in \mathbb{R}$ is symmetric under $\nu \leftrightarrow -\nu - 1$ so that, without loss of generality, we may take $\nu \geq -1/2$ and $\nu\left(\nu + 1\right) \geq -1/4$. If further $\nu \in \mathbb{N}$, then the first $2\nu + 1$ Lam\'{e} functions are polynomials (Lam\'{e} polynomials), while the remaining, transcendental solutions \textit{coexist}, i.e.,
\begin{IEEEeqnarray}{ll}
a_\nu^n = b_\nu^n \text{, for n, } \nu \in \mathbb{N} \text{ and } n \geq \nu + 1.
\end{IEEEeqnarray}
These results are nicely summarized in the following theorem \cite{MagnusWinkler66}: \\
\begin{theorem}
Lam\'{e}'s equation \ref{LameEquationIII}, displays \textit{coexistence} iff \, $\nu \in \mathbb{Z}$. It has exactly $\nu + 1$ instabilities, if $\nu \in \mathbb{N}$, and exactly $|\nu|$ instabilities, if $\nu \in \mathbb{Z}^-$. \\
\end{theorem}
The stability intervals in this case are \cite{NISTHandbook10}:\\
\begin{equation}
(a_\nu^0 \,,\, b_\nu^1) \; \cup \; (a_\nu^1 \,,\, b_\nu^2) \; \cup \; (a_\nu^2 \,,\, b_\nu^3) \; \cup \; \ldots \; \cup \; (a_\nu^{\nu - 1} \,,\, b_\nu^\nu) \; \cup \; (a_\nu^\nu \,,\, +\infty) \quad , \quad \nu \in \mathbb{N}. \\[12pt]
\end{equation}
Finally, we will say a few things about the Lam\'{e} functions of imaginary periods. We first observe that \ref{LameEquationIII} has the following symmetry \cite{Erdelyietal55,  NISTHandbook10, Dunne02, DunneShifman02}:
\begin{IEEEeqnarray}{c}
u' = i \left(u - \mathbb{K}\left(k^2\right) - i \, \mathbb{K}'\left(k^2\right)\right) \nonumber \\[6pt]
h' = \nu\left(\nu + 1\right) - h \quad , \quad k'\,^2 = 1 - k^2, \label{DualityLameTransform}
\end{IEEEeqnarray}
so that, when $z\left(u\right)$ has a real period of $2\,p\,\mathbb{K}$ ($p = 1,2$) and satisfies \ref{LameEquationIII}, $z'\left(u'\right) \equiv z\left(u\right)$ will have an imaginary period $2\,i\,p\,\mathbb{K}$ and will satisfy the transformed equation: \\
\begin{IEEEeqnarray}{l}
\frac{d^2z}{du'^2} + \left[h' - \nu\left(\nu + 1\right) k'\,^2 sn^2\left(u' | k'\,^2\right)\right] z = 0. \label{LameEquationIV}
\end{IEEEeqnarray} \\
It turns out that the duality \ref{DualityLameTransform}, interchanges the bands of stability with the gaps of instability, in \ref{LameBands} \cite{Dunne02, DunneShifman02}\nocite{Ince26, WhittakerWatson27}.
\normalsize
\bibliographystyle{JHEP}
\bibliography{Bibliography}
\end{document}